\documentclass[useAMS,usenatbib,letterpaper]{mnras}
\usepackage{epsfig,amsmath,amssymb,natbib}
\usepackage{aas_macros}

\begin{document}

\title[Empirical Ising Galaxy Bias]{\LARGE Empirical Validation of the Ising Galaxy Bias Model}

\author[A. Repp \& I. Szapudi]{Andrew Repp\ \& Istv\'an Szapudi\\Institute for Astronomy, University of Hawaii, 2680 Woodlawn Drive, Honolulu, HI 96822, USA}

\date{\today; to be submitted to MNRAS}

\label{firstpage}
\pagerange{\pageref{firstpage}--\pageref{lastpage}}
\maketitle

\begin{abstract}
\citet{Ising1} present a physically-motivated galaxy bias model which remains physical in low-density regions and which also provides a better fit to simulation data than do typical survey-analysis bias models. Given plausible simplifying assumptions, the physics of this model (surprisingly) proves to be analogous to the Ising model of statistical mechanics. In the present work we present a method of testing this Ising bias model against empirical galaxy survey data. Using this method, we compare our model (as well as three reference models -- linear, quadratic, and logarithmic) to SDSS, 6dFGS, and COSMOS2015 results, finding that for spectroscopic redshift surveys, the Ising bias model provides a superior fit compared to the reference models. Photometric redshifts, on the other hand, introduce enough error into the radial coordinate that none of the models yields a good fit. A physically meaningful galaxy bias model is necessary for optimal extraction of cosmological information from dense galaxy surveys such as \emph{Euclid} and WFIRST. 
\end{abstract}


\section{Introduction}
\label{sec:intro}
Large-scale galaxy surveys in the near future (e.g., \emph{Euclid}, \citealp{Euclid}; and WFIRST, \citealp{WFIRST}) will return data encoding a wealth of cosmological information. In particular, this information is expected to be invaluable in constraining theories of dark energy and modified gravity.

However, these surveys almost by definition use galaxy counts as a proxy for the underlying dark matter density, and it is known that galaxies are biased tracers of mass which typically cluster more strongly than dark matter (see, e.g., \citealp{Kaiser1984,Bardeen1986}). Therefore, in order to constrain cosmology with galaxy counts, one must model the relationship between the matter overdensity (defined as $\delta = \rho/\overline{\rho} - 1$, where $\rho$ is the matter density) and the galaxy overdensity (defined as $\delta_g = N/\overline{N} - 1$, where $N$ is the galaxy count in a survey cell). Although detailed bias models are of great theoretical importance, at the moment they are not practical for survey data analysis due to their large number of nuisance parameters.

The most common bias model for survey data analysis is the linear model (see e.g., \citealp{Hoffmann2017}) which expands the matter-galaxy relationship to first order ($\delta_g = b \delta$). One can account for mild non-linearities by including a quadratic term ($\delta_g = b_1 \delta + b_2\delta^2/2$), although surveys such as DES (Dark Energy Survey -- see \citealp{DESResults}) have not found it worthwhile to proceed past linear order. In both cases, however, one expects the linear expansion parameter ($b$ or $b_1$) to exceed unity in order to properly model the enhanced clustering of galaxy counts with respect to dark matter. For this reason the linear and quadratic models can readily yield $\delta_g < -1$ (an unphysical result, predicting negative galaxy counts) when applied to voids. One can preserve physicality in voids by working to linear order in log space (logarithmic bias, $\ln(1+\delta_g) = b \ln(1+\delta)$; see, e.g., \citealp{delaTorre2013}), but this procedure sacrifices the accuracy of the fit in high-density regions (see \citealp{Ising1} and the remainder of this work). 

The unphysicality of the linear and quadratic models in voids represents a significant shortcoming, given that voids at non-linear scales can contain up to half of the information inherent in survey results (e.g., \citealp{WCS2015Forecast}). In addition, the screening effects required by theories of modified gravity are detectable only in these voids. Thus, on the one hand, the standard linear approach is quite reasonable on larger (linear) scales where $\delta \sim 0$; in particular, the linear bias model emerges naturally from linear perturbation theory \citep{Desjacques2018}. At smaller scales, on the other hand, a linear bias produces unphysical results and does not reproduce the exponential decline noted in voids by \citep{Neyrinck2014}. Thus the most widespread bias model for data analysis fails in regions which contain significant amounts of valuable information regarding dark energy and modified gravity. Hence, effective utilization of the data from upcoming surveys requires an alternate model which makes accurate predictions for low-density regions (voids) as well as high-density regions (clusters).

Potential alternatives for analyzing the dark matter-galaxy relationship include full-scale halo modeling and/or hydrodynamical simulations. However, the former is difficult and requires multiple nuisance parameters, whereas the latter suffers from uncertainty regarding significant aspects of galaxy formation and evolution. It also requires modeling of a vast range of scales, typically introducing simplifications of unknown impact to render the calculations feasible.

In practice, extraction of cosmological information from galaxy surveys requires marginalization over the bias parameters(s). Hence a practical galaxy bias model requires (a) reasonable predictions for both high and low density regions -- thus, in particular, physical predictions for voids; and (b) comparability in terms of free parameters to the models typically employed in cosmological inference. This approach seeks a physical model with a minimal number of parameters for use in survey data analysis. Likewise, we here consider local bias models only, in order to minimize the model complexity.

For these reasons \citet{Ising1} present an Ising galaxy bias model. Ising models (first applied to ferromagnetism) employ two thermodynamically-distinct states to characterize populations; our model applies this formalism to the population of dark matter subhaloes, which can be in either a galaxy-hosting state or a non-hosting state. In this context, a simple set of physical assumptions (listed in Section~\ref{sec:models} of this work) yields an Ising model predicting a mean number of galaxies which, in dense regions, is proportional to the matter density and, in voids, drops exponentially to zero. The Ising model incorporates stochasticity, and (like the quadratic model) it contains only two free parameters. (See Section~\ref{sec:models} of this paper for an overview).

\citet{Ising1} then compare the Ising model (as well as the linear, quadratic, and logarithmic models) to the Millennium Simulation and a corresponding galaxy catalog; they find that at small scales ($\la 10h^{-1}$Mpc) the Ising model is vastly superior to the others, while at large scales ($\ga 30 h^{-1}$Mpc, where conventional models are already known to be adequate) the linear and quadratic models are preferable.

However, comparison with simulations is no substitute for comparison to actual galaxy survey results. These surveys typically do not provide access to the underlying dark matter content of survey cells; thus we cannot perform a direct comparison between matter and galaxy densities and must devise an alternate means of evaluating bias models. In this work we present a method for discriminating among bias models on the basis of empirical galaxy counts; we then apply this method to three galaxy survey data sets.

We organize this work as follows: first, Section~\ref{sec:models} provides an overview of the galaxy bias models considered in this work. Section~\ref{sec:PN} develops the method for comparison of the bias models to data and validates this method with simulations. Section~\ref{sec:data} performs the comparison to three galaxy survey data sets (namely, galaxies from the SDSS Main Galaxy Sample, the 6dFGS survey, and COSMOS2015); discussion and conclusion follow in Sections~\ref{sec:disc} and \ref{sec:concl}. Unless otherwise noted, we assume a concordance cosmology in which $h=0.7$, $\Omega_m=0.3$, and $\Omega_\Lambda=0.7$.

\section{Galaxy Bias Models}
\label{sec:models}

Galaxy bias models parametrize the relationship between the matter density contrast ($\delta = \rho/\overline{\rho} - 1$, where $\rho$ is matter density in a survey cell) and the galaxy number density contrast ($\delta_g = N/\overline{N} - 1$, where $N$ is the number of galaxies in a survey cell). In this paper we consider three reference models for galaxy bias. The first (linear bias) is the only model commonly employed in data analysis. The second and third (quadratic and logarithmic) represent some of the simplest generalizations of the linear model, the quadratic expanding the relationship to second order and the logarithmic guaranteeing physicality by working in log space. Mathematically, the three reference models are as follows:
\begin{align}
\delta_g & = b \delta \hspace{3mm}\mbox{(linear bias)}\label{eq:linbias}\\
\delta_g & = b_1 \delta + \frac{b_2}{2}\delta^2 \hspace{3mm}\mbox{(quadratic bias)}\\
\ln(1+\delta_g) & = b \ln(1 + \delta) \hspace{3mm}\mbox{(logarithmic bias)}\label{eq:logbias}
\end{align}
To model the enhanced clustering of galaxies (with respect to dark matter), typical values of $b$ (and $b_1$) must exceed unity, and thus the linear and quadratic models can yield unphysical results ($\delta_g < -1$) in low-density regions.

In contrast, the Ising model begins with the following  physical assumptions: (1) the formation of a galaxy in a given dark matter subhalo depends (to first order) upon initial conditions and local physics only; (2) one can treat the subhaloes hosting a given type of galaxy as roughly equivalent; (3) the formation of clustered galaxies is energetically favorable; and (4) the number of subhaloes in a survey cell is (on average) proportional to the cell's matter density. From these assumptions, \citet{Ising1} derive the following expression for $M$, defined as the mean number of galaxies per mass:
\begin{equation}
\label{eq:final_FD}
 M \equiv \langle N \rangle_A \cdot (1 + \delta)^{-1} = \frac{b_I\overline{N}}{1 + \exp\left(\frac{A_t-A}{T}\right)}.
\end{equation}
Here $A \equiv \ln(1+\delta)$ as a proxy for initial overdensity \citep{CarronSzapudi2013}, which pursuant to our first physical assumption determines the mean number of galaxies $M$ per unit mass. Equation~\ref{eq:final_FD} also includes the three parameters $A_t$, $T$, and $b_I$. Hence for a given (log) matter density $A$ in a survey cell, the model predicts the expected number of galaxies $\langle N \rangle_A = M(A) e^A$ in that cell. This formalism thus incorporates stochasticity into the model. We likewise interpret the three reference models stochastically (i.e., as predicting $\langle \delta_g \rangle_\delta$ rather than $\delta_g$ itself), thus employing them in a somewhat more sophisticated manner than is typical.

Equation~\ref{eq:final_FD} is formally a Fermi-Dirac function with a transition density $A_t$ and asymptotic values $M=0$ (for $A \ll A_t$) and $M=b_I\overline{N}$ (for $A \gg A_t$). Thus in high-density regions, $\langle N \rangle_A$ approaches $b_I\overline{N}(1+\delta)$, so that the number of galaxies is proportional to the amount of underlying matter, as in the linear model; for low-density regions, the number of galaxies drops exponentially to zero, as \citet{Neyrinck2014} observe. Of the remaining parameters, $b_I$ is analogous to the linear bias parameter in high-density regions, and $T$ parametrizes the suddenness of the transition from the low- to high-density regimes.

We also note the existence of the net-galaxy constraint
\begin{equation}
b_I = \left( \int dA\,\, \mathcal{P}(A) \frac{e^A}{1 + \exp\left(\frac{A_t-A}{T}\right)} \right)^{-1},
\label{eq:constraint}
\end{equation}
where $\mathcal{P}(A)$ is the (log) dark matter probability distribution. This constraint restricts parameter values to ensure that the total number of galaxies matches the survey results, and as a result the Ising model requires only two free parameters (as does the quadratic). In this work we use the GEV prescription for $\mathcal{P}(A)$ from \citet{ReppApdf} (though see the discussion in Section~\ref{sec:PN}).

\citet{Ising1} also investigate the utility of modifying the Ising model to better match the observed first-order behavior in high-density regions. Doing so requires an additional free parameter $k$, with
\begin{equation}
M = \frac{b_I\overline{N}}{1 + \exp\left(\frac{A_t-A}{T}\right)} + \left(1-b_I\right)\overline{N}e^{-A} \exp\left(-ke^{-A}\right).
\label{eq:modIsing}
\end{equation}
In this model, the net-galaxy constraint of Equation~\ref{eq:constraint} becomes
\begin{equation}
b_I = \frac{1-I_2}{I_1-I_2},
\label{eq:modIsconstraint}
\end{equation}
where $I_1$ and $I_2$ are the integrals
\begin{align}
I_1 & = \int dA\,\, \mathcal{P}(A) \frac{e^A}{1 + \exp\left(\frac{A_t-A}{T}\right)}\label{eq:defI1}\\
I_2 & = \int dA\,\, \mathcal{P}(A) \exp\left(-ke^{-A}\right)\label{eq:defI2}.
\end{align}
In most cases this modified Ising model improves the fit (as measured by the $\chi^2$ per degree of freedom), at the cost of an additional parameter. In the following sections, we will test this modified Ising model alongside the (unmodified) Ising model and the reference models (linear, quadratic, and logarithmic).

Before proceding to these tests, it will be useful to write Equations~\ref{eq:linbias}--\ref{eq:logbias} in terms of $A$, $\overline{N}$, and $M\equiv \langle N \rangle_A \cdot (1+\delta)^{-1}$, in order to make them fully analogous to Equation~\ref{eq:final_FD}:
\begin{align}
M & = \overline{N} \left(b + (1-b)e^{-A} \right) \hspace{3mm}\mbox{(linear bias)}\label{eq:lin}\\
M & = \overline{N} \left( \frac{b_2}{2}e^A + \left(b_1-b_2\right) + e^{-A}\left( \frac{b_2}{2} - b_1 + 1 \right)\right)\label{eq:quad}\\
  &   \hspace{5cm} \mbox{(quadratic bias)}\nonumber\\
M & = \overline{N} e^{A(b-1)}\hspace{3mm}\mbox{(logarithmic bias)}\label{eq:log}
\end{align}
It is Equations~\ref{eq:final_FD}, \ref{eq:modIsing}, and \ref{eq:lin}--\ref{eq:log} which we shall fit to simulations and observations in the subsequent sections.

\section{Comparison Methods}
\label{sec:PN}
\citet{Ising1} compare various bias models to the results of the Millennium Simulation \citep{Springel2005} at $z=0$, 0.51, 0.99, and 2.07. Using the corresponding galaxy catalog of \citet{Bertone2007}, they apply a stellar mass cut of $M_\star \geq 10^9h^{-1}M_\odot$, perform counts-in-cells at scales from $1.95h^{-1}$ Mpc to $31.25h^{-1}$ Mpc, and find the best fits for Equations~\ref{eq:final_FD} and \ref{eq:lin}--\ref{eq:log}. The results demonstrate that at small scales ($\la 10h^{-1}$ Mpc) the Ising model is vastly superior to the reference models, whereas at large scales ($\ga 30h^{-1}$ Mpc) the linear and quadratic models fare somewhat better. Note that this large-scale regime is precisely where the linear model is already known to be adequate.

However, this method of analysis (i.e., direct comparison of matter and galaxy overdensities) is inapplicable to typical galaxy surveys, which use galaxies as tracers of mass and thus have no access to the dark matter content of each survey cell. For this reason, in the current section we develop a procedure for evaluating galaxy bias models by comparison to the observed galaxy probability distribution $\mathcal{P}(N)$.

\subsection{Modeling $\mathcal{P}(N)$}
\label{sec:PNmethod}
\citet{Ising1} report an initial attempt at modeling $\mathcal{P}(N)$ (where $N$ is the number of galaxies in a given survey cell) by ignoring the scatter of $N$ about $\langle N \rangle_A$ and matching the cumulative distribution functions of $A$ (derived from \citealt{ReppApdf}) and $N$ (from counting galaxies). We note that the results of this abundance-matching procedure provide qualitative information about the number of galaxies per dark matter mass and in particular reproduce the transition between low- and high-density regimes. The abundance-matching technique is thus useful, but it proves insufficiently accurate for meaningful model fits.

Thus, in order to compare the predictions of the bias models to empirical galaxy counts, we model the distribution $\mathcal{P}(N)$ as follows. Given the theoretical (GEV) distribution $\mathcal{P}(A)$ of \citet{ReppApdf} and assuming a Poisson distribution for the conditional probability $P(N|A)$, we can write
\begin{align}
\mathcal{P}(N) & = \int dA \,\,\mathrm{Pois}\left(N;\langle N \rangle_A \right)\,\mathcal{P}(A)
\label{eq:Nprobs}\\
  & = \int dA \,\, \mathrm{Pois}\left(N; M(A)e^A \right)\, \mathcal{P}(A),
\label{eq:probs}
\end{align}
where $\langle N \rangle_A \equiv M(A)e^A$ in Equations~\ref{eq:final_FD}, \ref{eq:modIsing}, and \ref{eq:lin}--\ref{eq:log}. Inversion of Equation~\ref{eq:probs} would then yield the values of $M(A)$ (or of $\langle N \rangle_A$) in terms of measured galaxy counts.

Unfortunately, such inversion is non-trivial, and the non-linearities of the functions involved render numerical approaches unstable \citep{SzapudiPan2004}. Thus we employ instead a forward-modeling approach and attempt to determine the parameters for each model which best reproduce the observed distribution of $\mathcal{P}(N)$.

In order to apply Equation~\ref{eq:probs} to simulated or empirical data, we must navigate two complications. First, we must determine a consistent method of handling non-physical predictions of the reference models; in particular, Equations~\ref{eq:lin} and \ref{eq:quad} can predict negative values for $\langle N \rangle_A$, in which case the Poisson distribution of Equation~\ref{eq:Nprobs} is undefined. One option would be to arbitrarily set $\mathcal{P}(N|A)=0$ in such cases; however, this procedure would artificially introduce a lower bound into the model. The result would be an artificially-sigmoid (and thus artificially Ising-like) function, obscuring the differences between the models. (The most serious drawback of the linear and quadratic models is their unphysicality, so a procedure which artificially removes that unphysicality would yield an unfair comparison.) As an alternative, we choose to consider a prediction of $\langle N \rangle_A < 0$ as implying a negative number of galaxies in the survey cell, so that $\sum_{N=0}^\infty \mathcal{P}(N) < 1$. This unphysical result is in keeping with the unphysical nature of the prediction.

\begin{figure*}
\leavevmode\epsfxsize=16cm\epsfbox{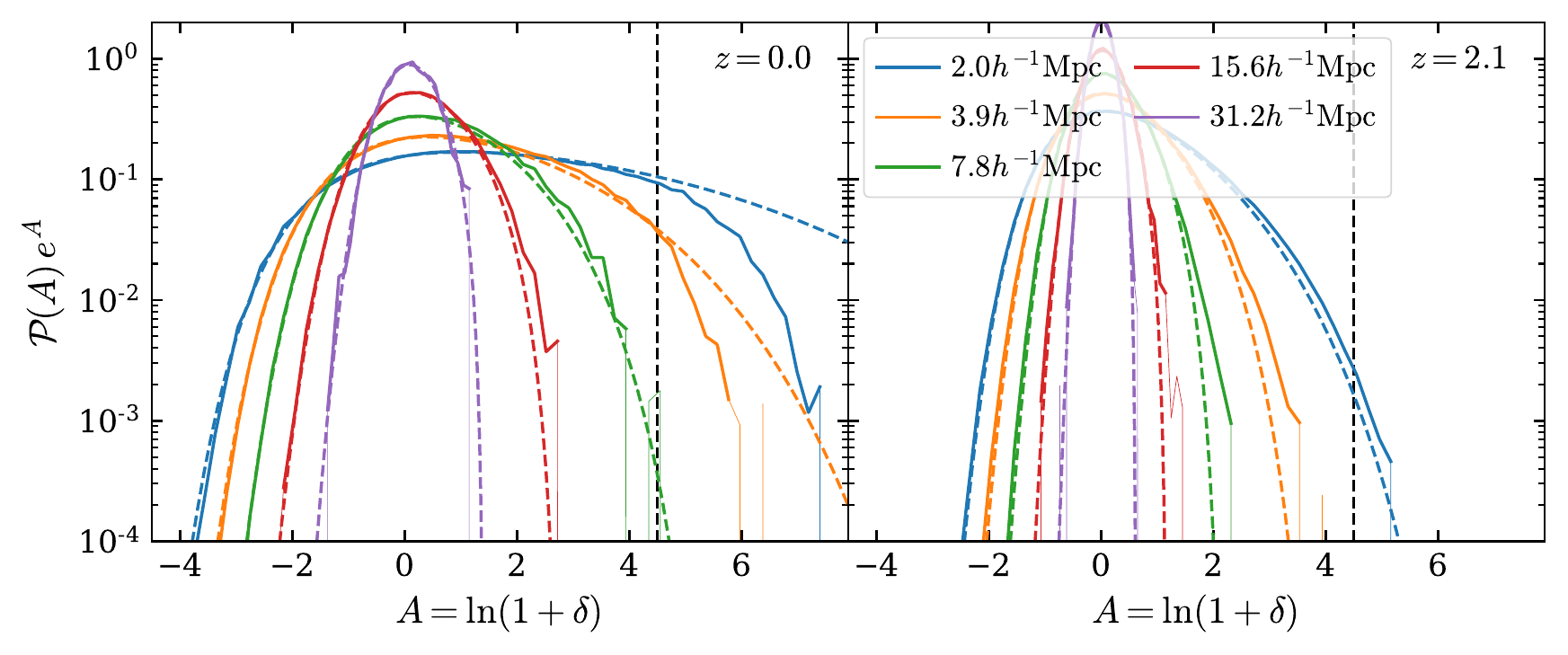}
\caption{Exponentially-enhanced probability distributions $\mathcal{P}(A) \exp(A)$ for the log matter density $A \equiv \ln(1+\delta)$ for various smoothing scales. Solid lines show the probability measured from the Millennium Simulation, with thin lines showing the transition to the Poisson noise regime (0 or 1 cells in the $A$-bin). Dashed curves show the distribution predicted by the GEV model of \citet{ReppApdf}. The left and right panels show results from $z=0$ and 2.1, respectively. The vertical dashed lines show the location of $A_\mathrm{cut}$ discussed in the text. In addition to the exponential cutoff considered in the text, one notes a slight probability excess at high $A$-values. Because this excess becomes significant only at low probabilities, it has little impact on the fits performed in this work.}
\label{fig:PA_compare}
\end{figure*}

The second complication is the limited accuracy of the Generalized Extreme Value (GEV) prediction at high values of $A$ for the exponentially enhanced distribution $\mathcal{P}(A)e^A$. \citet{ReppApdf} show that this GEV prescription for $\mathcal{P}(A)$ itself yields a cumulative distribution function with an accuracy better than 2 per cent. For very large $A$-values, however, the GEV prescription tends to overpredict the matter density (as noted by \citealp{Klypin2018}); furthermore, multiplication by an exponential (to apply the net-galaxy constraint of Equation~\ref{eq:constraint}) exacerbates the inaccuracy for small scales (see Fig.~\ref{fig:PA_compare}).

Circumventing this difficulty requires an estimate of the cutoff value $A_\mathrm{cut}$ beyond which we cannot safely use the GEV prescription. It is tempting to speculate that the downturn in the measured values of $\mathcal{P}(A)$ reflects the onset of virialization at $\delta_\mathrm{vir} = 177$. If we conservatively take half of this value, we obtain $A_\mathrm{cut} = \ln(1+\delta_\mathrm{vir}/2) = 4.5$. Reference to Fig.~\ref{fig:PA_compare} shows that this value for $A_\mathrm{cut}$ reasonably approximates the scale at which the dropoff begins, at least for smoothing scales which reach this value before experiencing significant shot noise. (Note that this cutoff is not the only inaccuracy observable in Fig.~\ref{fig:PA_compare} or discussed in \citealp{ReppApdf}, but it is the most significant for our forward-modeling process.) Thus, whether or not the dropoff is indeed due to virialization, we adopt $A_\mathrm{cut} = 4.5$ for the following analysis. We also note that the survey data which we analyse in Sections~\ref{sec:6dFGS}--\ref{sec:COSMOS} never exceed this density, which is therefore relevant only to the validation in this section of this method.

We now require versions of Equations~\ref{eq:constraint} and \ref{eq:modIsconstraint} which consider only values $A < A_\mathrm{cut}$. In Appendix~\ref{app:cut} we show that required equations are as follows. For the (non-modified) Ising model, the net-galaxy constraint (analogous to Equation~\ref{eq:constraint}) is
\begin{equation}
b_I = f/I_1^\mathrm{cut},
\label{eq:bigeq_unmod}
\end{equation}
where
\begin{equation}
f = \frac{1}{N_\mathrm{cells}\overline{N}}\sum_{N=0}^{N_\mathrm{cut}} N\cdot n(N)
\label{eq:deff}
\end{equation}
and
\begin{equation}
I_1^\mathrm{cut} = \int_{-\infty}^{A_\mathrm{cut}} dA\,\mathcal{P}(A) \frac{e^A}{1+\exp\left(\frac{A_t-A}{T}\right)}.
\label{eq:defI1cut}
\end{equation}
Here $n(N)$ is the number of survey cells containing $N$ galaxies, $N_\mathrm{cells}$ is the total number of survey cells, and $N_\mathrm{cut} = b_I \overline{N} \exp\left(A_\mathrm{cut}\right) = 90 b_I \overline{N}$. Equation~\ref{eq:constraint} (to which Equations~\ref{eq:bigeq_unmod}--\ref{eq:defI1cut} are analogous) requires empirical determination of $\overline{N}$; Equation~\ref{eq:deff} requires in addition determination of the $n(N)$-values from the survey data results. Note that the appearance of $b_I$ on both sides of Equation~\ref{eq:bigeq_unmod} (on the right-hand side through $N_\mathrm{cut}$ in Equation~\ref{eq:deff}) requires an iterative numerical solution. It is clear that as $A_\mathrm{cut} \rightarrow \infty$, we have $f \rightarrow 1$, and $I_1^\mathrm{cut} \rightarrow I_1$ of Equation~\ref{eq:defI1}, so that in this limit Equation~\ref{eq:bigeq_unmod} reduces to Equation~\ref{eq:constraint}.

For the modified Ising model, we have a similar result, namely, a net-galaxy constraint of 
\begin{equation}
b_I = \frac{f-I_2^\mathrm{cut}}{I_1^\mathrm{cut} - I_2^\mathrm{cut}},
\label{eq:bigeqmod}
\end{equation}
where
\begin{equation}
I_2^\mathrm{cut} = \int_{-\infty}^{A_\mathrm{cut}} dA\,\mathcal{P}(A) \exp\left(-ke^{-A}\right).
\label{eq:defI2cut}
\end{equation}
Equation~\ref{eq:bigeqmod} is the analog of Equation~\ref{eq:modIsconstraint}, and in the limit $A_\mathrm{cut} \rightarrow \infty$ we have $I_2^\mathrm{cut} \rightarrow I_2$ of Equation~\ref{eq:defI2}.

Finally, for each of the reference models (linear, quadratic, and logarithmic) we can write $N_\mathrm{cut} = M(A_\mathrm{cut})\exp(A_\mathrm{cut})$, with $M(A)$ given by Equations~\ref{eq:linbias}--\ref{eq:logbias}.

In summary, we now have a practical method for comparing each bias model to the observed distribution of galaxy counts. This method models $\mathcal{P}(N)$ using the theoretical GEV prescription for $\mathcal{P}(A)$ and a Poisson distribution for $\mathcal{P}(N|A)$, with $\langle N \rangle_A$ determined by the galaxy-bias model under evaluation. The inaccuracy of the GEV prescription at high $A$-values invites the following modifications. First, in comparing the distribution $\mathcal{P}(N)$ observed in a survey to that predicted by the various bias models, we will cut off the reduced-$\chi^2$ evaluation at the value(s) of $N_\mathrm{cut}$ applicable to each model. Second, in evaluating the Ising model (or the modified Ising model), we should use Equation~\ref{eq:bigeq_unmod} (or, respectively, Equation~\ref{eq:bigeqmod}) to impose the net-galaxy constraint.

\subsection{Validation of $\mathcal{P}(N)$ Model}
\label{sec:PNvalid}
We now validate this modeling procedure by comparison to the Millennium Simulation catalogs (described at the start of Section~\ref{sec:PN}) to see how well each galaxy bias prescription can reproduce the actual distribution $\mathcal{P}(N)$.

As a practical matter of determining probabilities, for higher $N$-values the number of cells containing $N$ galaxies is typically 1 or 0, resulting in significant shot noise. To obtain better limits on $\mathcal{P}(N)$ we thus bin the $N$-values when obtaining empirical probabilities. It turns out that the absolute values of $\chi^2_\nu$ are rather sensitive to the particular binning strategy employed, but the relative values at any given smoothing scale will still quantify the superiority of one model over another at that scale. The details of our binning strategy, as well as our procedure for estimating the uncertainties on $\mathcal{P}(N)$, appear in Appendix~\ref{app:PNerrs}. With these values, we can use the procedure of Section~\ref{sec:PNmethod} to determine the best-fitting parameter values -- and the corresponding best-fitting predicted $\mathcal{P}(N)$-distributions -- for each galaxy bias model. (Note that the procedure for estimating $\sigma^2_{\mathcal{P}(N)}$ ignores covariance between $N$-values, and thus some of the resultant $\chi^2_\nu$-values are significantly less than unity. Nevertheless the relative values of $\chi^2_\nu$ at a fixed smoothing scale and redshift quantify the different fitting capabilities of the bias models.) 
 
\begin{figure}
\leavevmode\epsfxsize=8cm\epsfbox{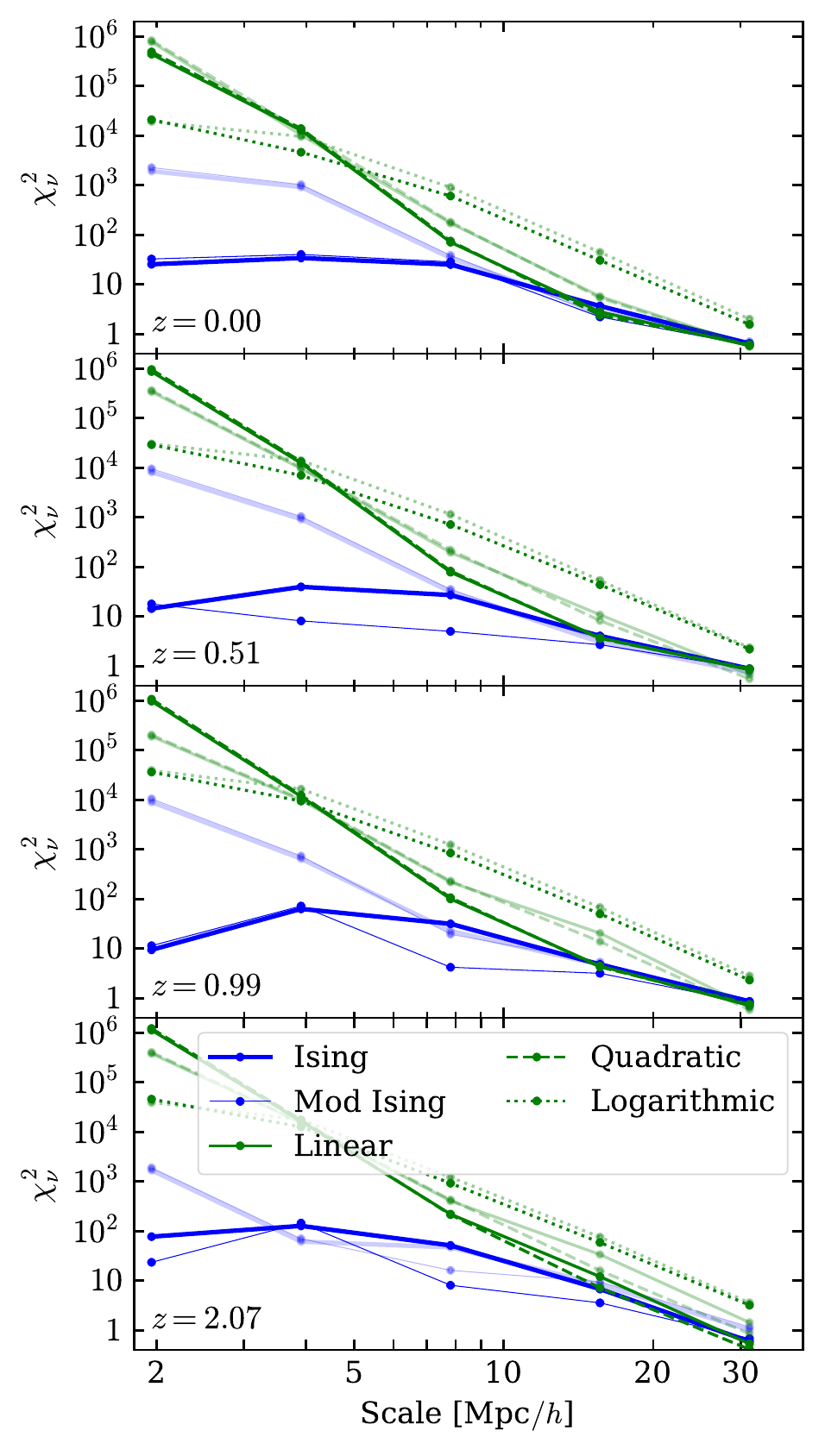}
\caption{$\chi^2$ per degree of freedom for the best fits of various models to the galaxy probability distribution $\mathcal{P}(N)$ of the Millennium Simulation catalog discussed in the text. Solid curves display the results of fits in real space, and semi-transparent curves display the results of fits in redshift space. The Ising model and its modified variant consistently outperform the standard models at small scales (most strikingly in real space, but also in redshift space) and yield comparable fits at large scales.}
\label{fig:chi2dof}
\end{figure}

\begin{figure*}
\leavevmode\epsfxsize=16cm\epsfbox{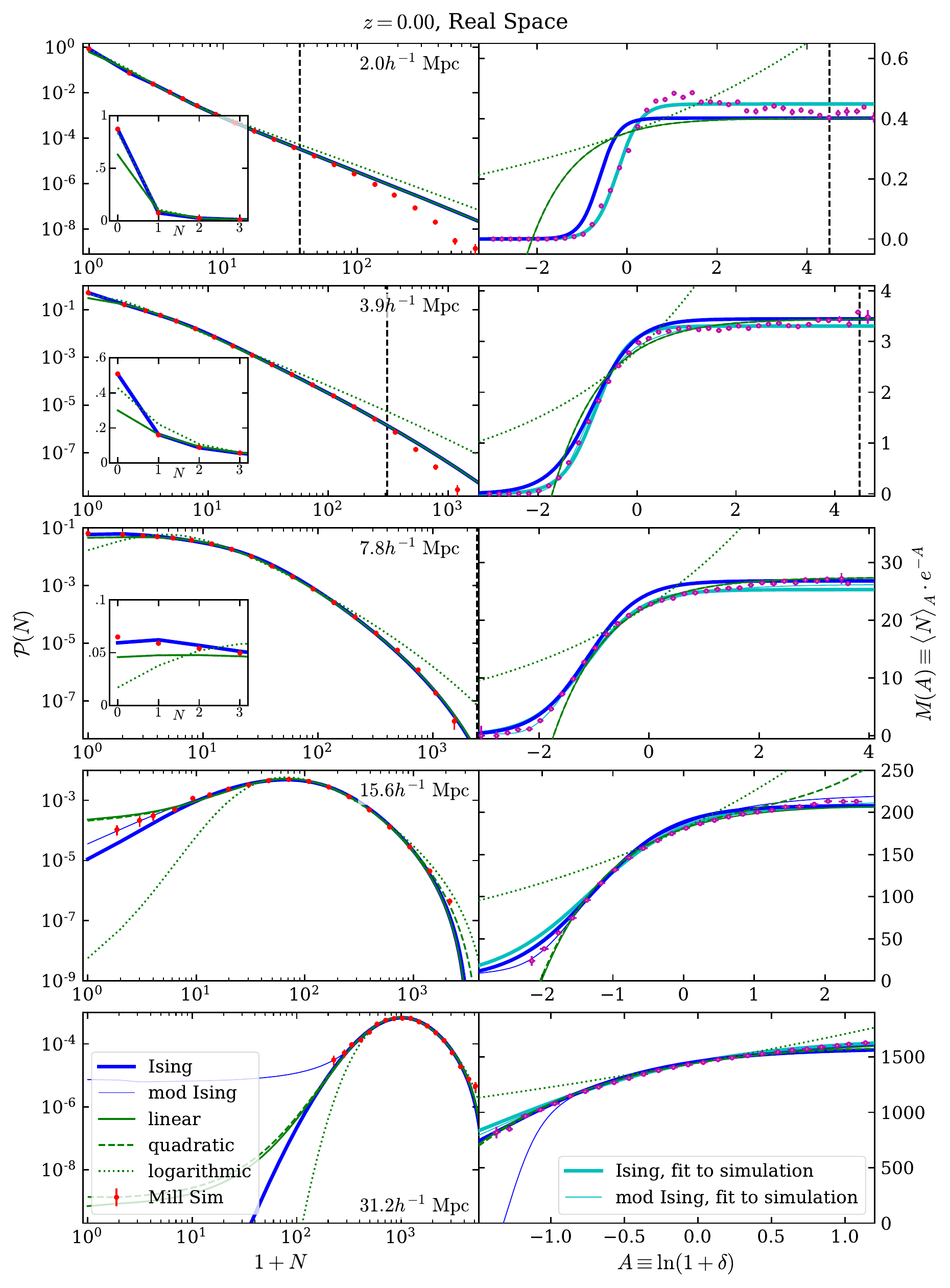}
\caption{Results of fitting bias models to Millennium Simulation galaxy distributions at $z = 0.$ Left panels: probability distributions $\mathcal{P}(N)$, where $N$ is the number of galaxies per survey cell, for various cell sizes. Red points show the probabilities observed in the Millennium Simulation galaxy catalog discussed in the text; blue and green curves show the predictions of five galaxy bias models. Insets show the low-$N$ distribution on a linear scale. Right panels: predictions of best-fitting models (from left-hand panels) for the average number of galaxies per unit mass ($M$ in Equation~\ref{eq:final_FD}) as a function of log matter density $A$. We also show (in cyan) the best-fitting Ising and modified-Ising models derived from direct comparison of $A$ and $N$ in \citet{Ising1} (see figs. 1--4 therein) and the Millennium Simulations values of $M(A)$ in magenta. Dashed vertical lines show the values of $N_\mathrm{cut}$ and $A_\mathrm{cut}$ adopted in Equations~\ref{eq:bigeq_unmod}--\ref{eq:defI2cut}.}
\label{fig:MS_PN0}
\end{figure*}

\begin{figure*}
\leavevmode\epsfxsize=16cm\epsfbox{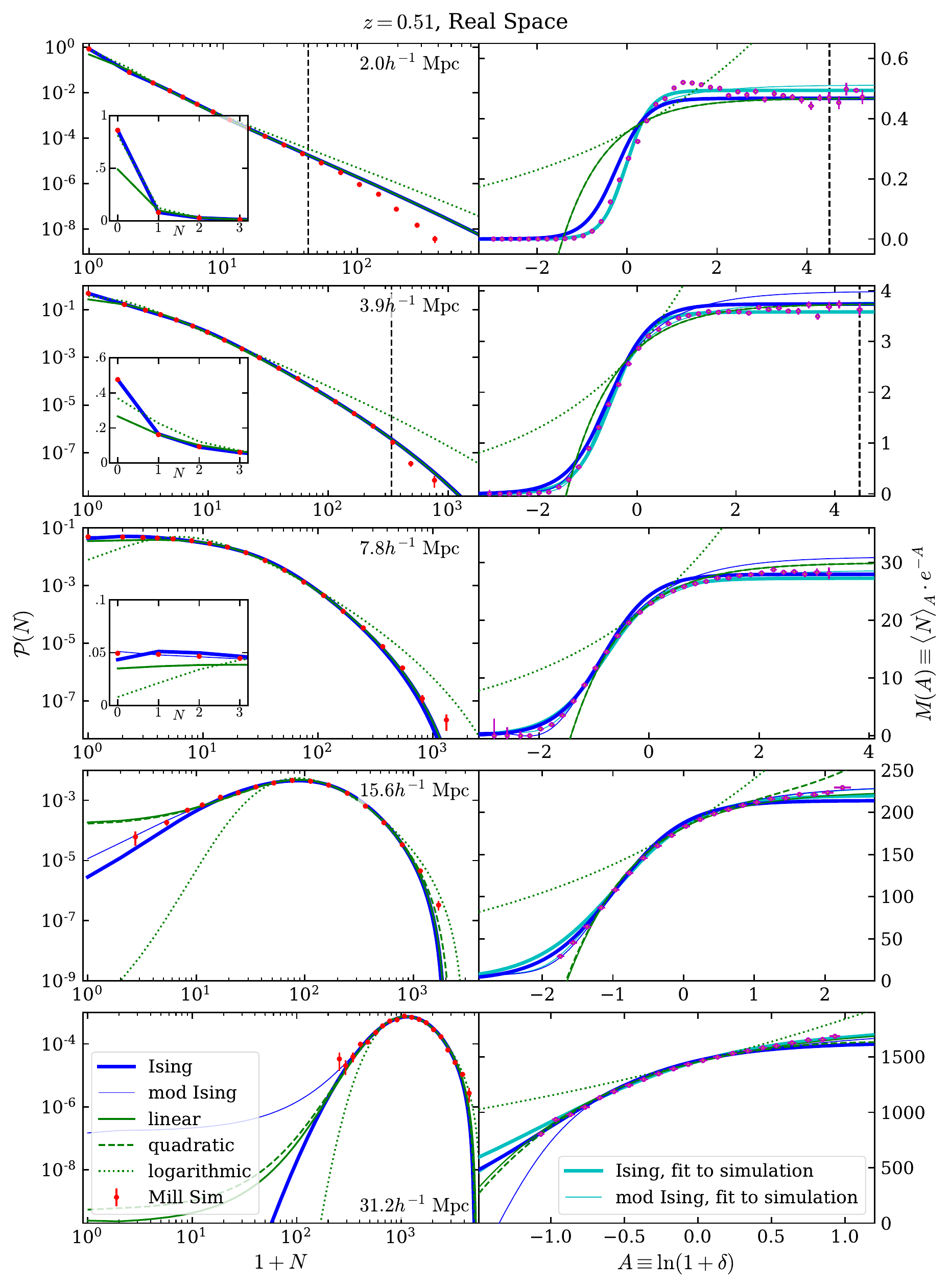}
\caption{Results of fitting bias models to Millennium Simulation galaxy distributions at $z = 0.51.$ See caption of Fig.~\ref{fig:MS_PN0} for description of symbols and colors.}
\label{fig:MS_PN51}
\end{figure*}

\begin{figure*}
\leavevmode\epsfxsize=16cm\epsfbox{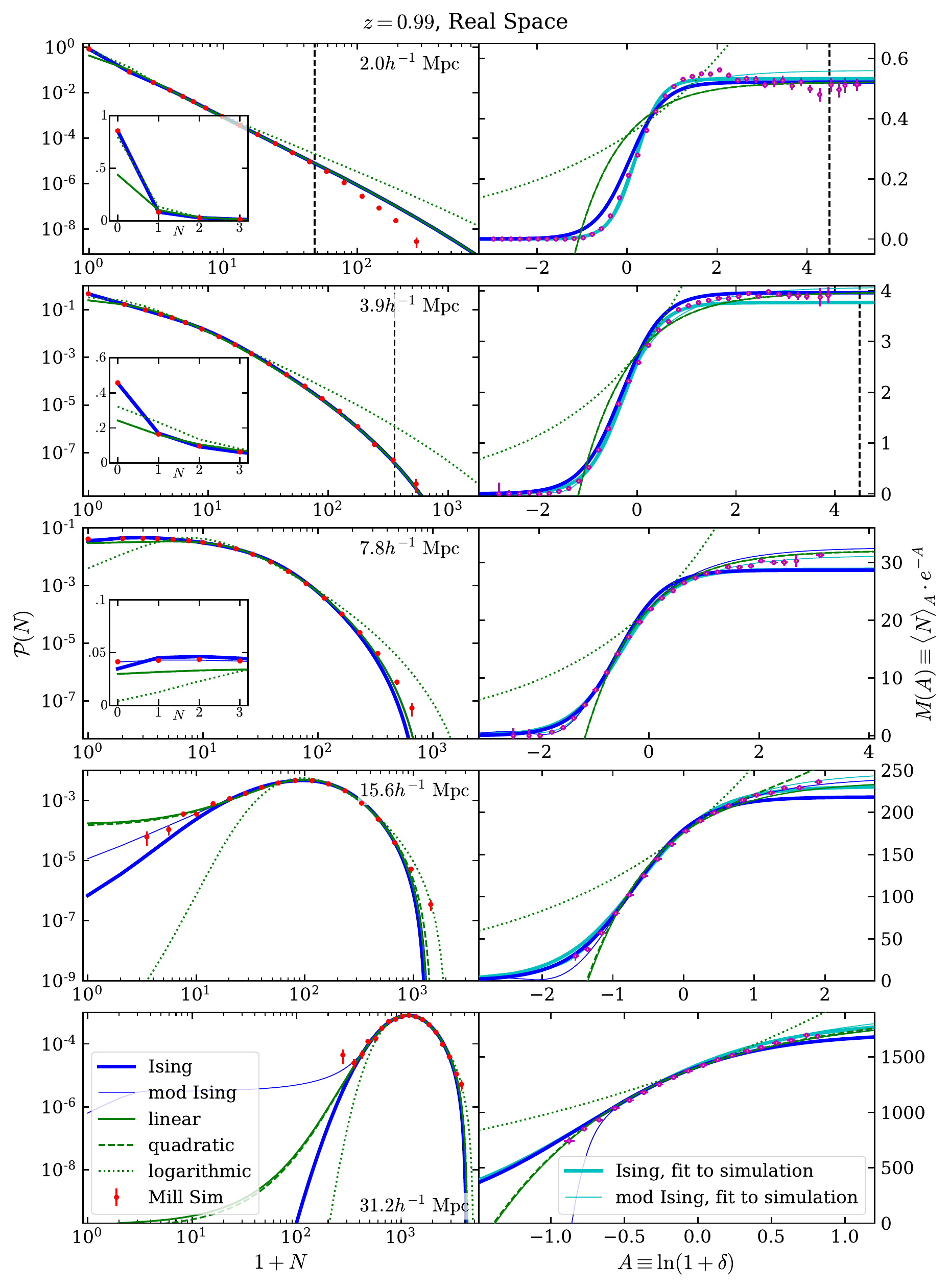}
\caption{Results of fitting bias models to Millennium Simulation galaxy distributions at $z = 0.99.$ See caption of Fig.~\ref{fig:MS_PN0} for description of symbols and colors.}
\label{fig:MS_PN99}
\end{figure*}

\begin{figure*}
\leavevmode\epsfxsize=16cm\epsfbox{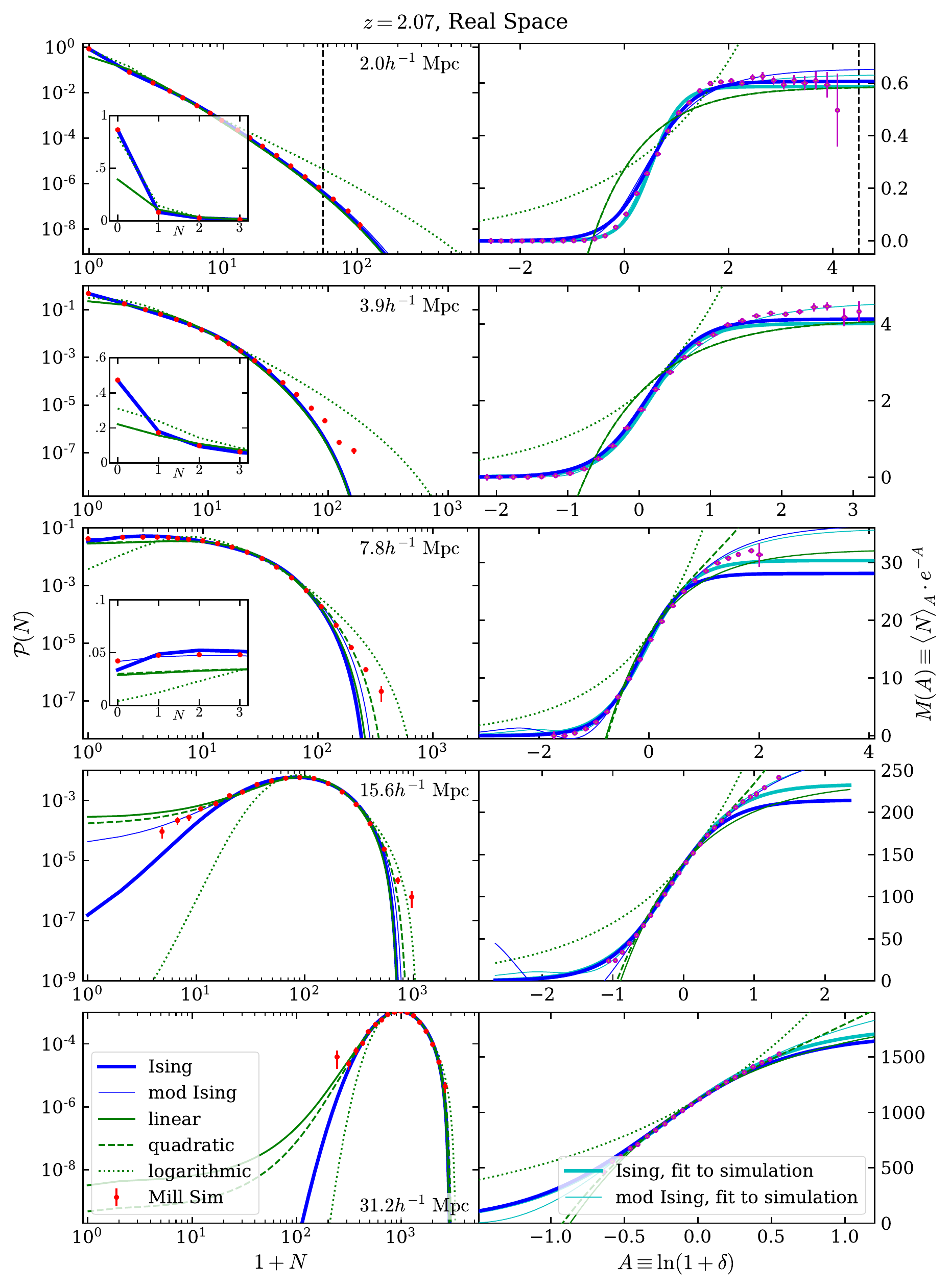}
\caption{Results of fitting bias models to Millennium Simulation galaxy distributions at $z = 2.07.$ See caption of Fig.~\ref{fig:MS_PN0} for description of symbols and colors.}
\label{fig:MS_PN2}
\end{figure*}

We perform these fits at the same redshifts and smoothing scales as \citet{Ising1}, namely, at redshifts $z = 0, 0.51, 0.99, 2.07$, and at five smoothing scales logarithmically-spaced from $1.95$--$31.25h^{-1}$Mpc. For each model we determine the best-fitting parameters and corresponding values of $\chi^2_\nu$ (i.e., the reduced $\chi^2$, or $\chi^2$ per degree of freedom). Fig.~\ref{fig:chi2dof} shows the resulting $\chi^2_\nu$ values for each model; the various fits themselves appear in Figs.~\ref{fig:MS_PN0}--\ref{fig:MS_PN2}, with the actual best-fitting parameter- and $\chi^2_\nu$-values appearing in Table~\ref{tab:MS_PN}. 

For each smoothing scale, Figs.~\ref{fig:MS_PN0}--\ref{fig:MS_PN2} show two panels. The left-hand panel displays the observed probabilities along with the best-fitting Ising, modified Ising, linear, quadratic, and logarithmic bias models. Note that though these panels use logarithmic scaling to better display low probabilities, the $\chi^2_\nu$ is most strongly affected by the low-$N$ fit, where the linear and quadratic models make unphysical predictions; to display this difference more clearly, we include insets (for the smallest three scales) showing $\mathcal{P}(N)$ for $N \le 3$ on a linear scale.

The right-hand panels of these figures show the predictions of these models for $M$ (from Equations~\ref{eq:final_FD}, \ref{eq:modIsing}, and \ref{eq:lin}--\ref{eq:log}) as a function of $A$. We also show (as cyan curves) the best-fitting Ising and modified-Ising model from the direct comparison (of galaxy and dark matter densities) reported in \citet{Ising1}; we also show relationship (magenta points) observed in the Millennium Simulation galaxy catalogs.

Let us consider the $\chi^2_\nu$-values of Fig.~\ref{fig:chi2dof}. The solid curves display the real-space fits of this section (for the semi-transparent curves, see Section~\ref{sec:zsp}). They reveal the same results noted in \citet{Ising1}, namely, a stark difference between the models at small scales -- with the Ising model vastly superior to the others -- fading to a slight preference for the linear and quadratic models once we reach linear scales. (Note that for surveys sparse enough to require analysis at these large scales, linear perturbation theory would in any case be sufficient, obviating the need for models such as the Ising.) We note in passing that these results suggest that it is hardly ever worthwhile to proceed from a linear to a quadratic model, as noted already by the Dark Energy Survey team \citep{DESResults}. Note also that we do not claim the Ising model to be a perfect representation even at smaller scales: the large values of $\chi^2_\nu$ make that fact clear, and it would indeed be surprising if the process of galaxy formation could be reduced to two or three parameters. But these results show, first, that the Ising model is vastly superior to the reference models at small scales; they show, second, that the modeling of $\mathcal{P}(N)$ described in this section effectively discriminates between bias models even without access to the underlying matter distribution values.

\subsection{Redshift-Space Distortion}
\label{sec:zsp}
Any three-dimensional analysis of empirical galaxy surveys must contend with the fact that they use redshift as a proxy for radial distance. As a result, departures of individual galaxies from the Hubble flow distort the calculated distances so that the redshift-space distribution of galaxies differs from the actual real-space distribution.

This distortion comprises two main components. First, on large scales matter is still falling toward overdensities (and evacuating voids). The result of this coherent motion is a flattening of structure (and elongation of voids) along the line of sight. \citet{Kaiser1987} provides the standard model for such distortions, in which the redshift-space volume element $d^3x_z$ replaces the real-space element $d^3x$. One would expect this transformation to have little impact on the bias-model fits because the distribution $\mathcal{P}(N)$ does not heavily depend on the survey cell shape. And while the distortion affects cell volume as well as shape, the effect is perturbative, and the fits of Section~\ref{sec:PNvalid} show that the Ising model is appropriate for a wide variety of cell volumes.

The second component of redshift-space distortion becomes important on small scales, where the thermal motion of virialized cluster members produces an apparent elongation (``Fingers of God'') of clusters along the line of sight. Consideration of typical intracluster velocities leads one to expect the greatest Finger-of-God impact to occur on scales below $\sim 5h^{-1}$Mpc. Thus we expect degradation of the Ising fit on these scales, but we also expect the Ising model (due to its physicality) to outperform the reference models even in redshift space at these scales. 

\begin{figure*}
\leavevmode\epsfxsize=16cm\epsfbox{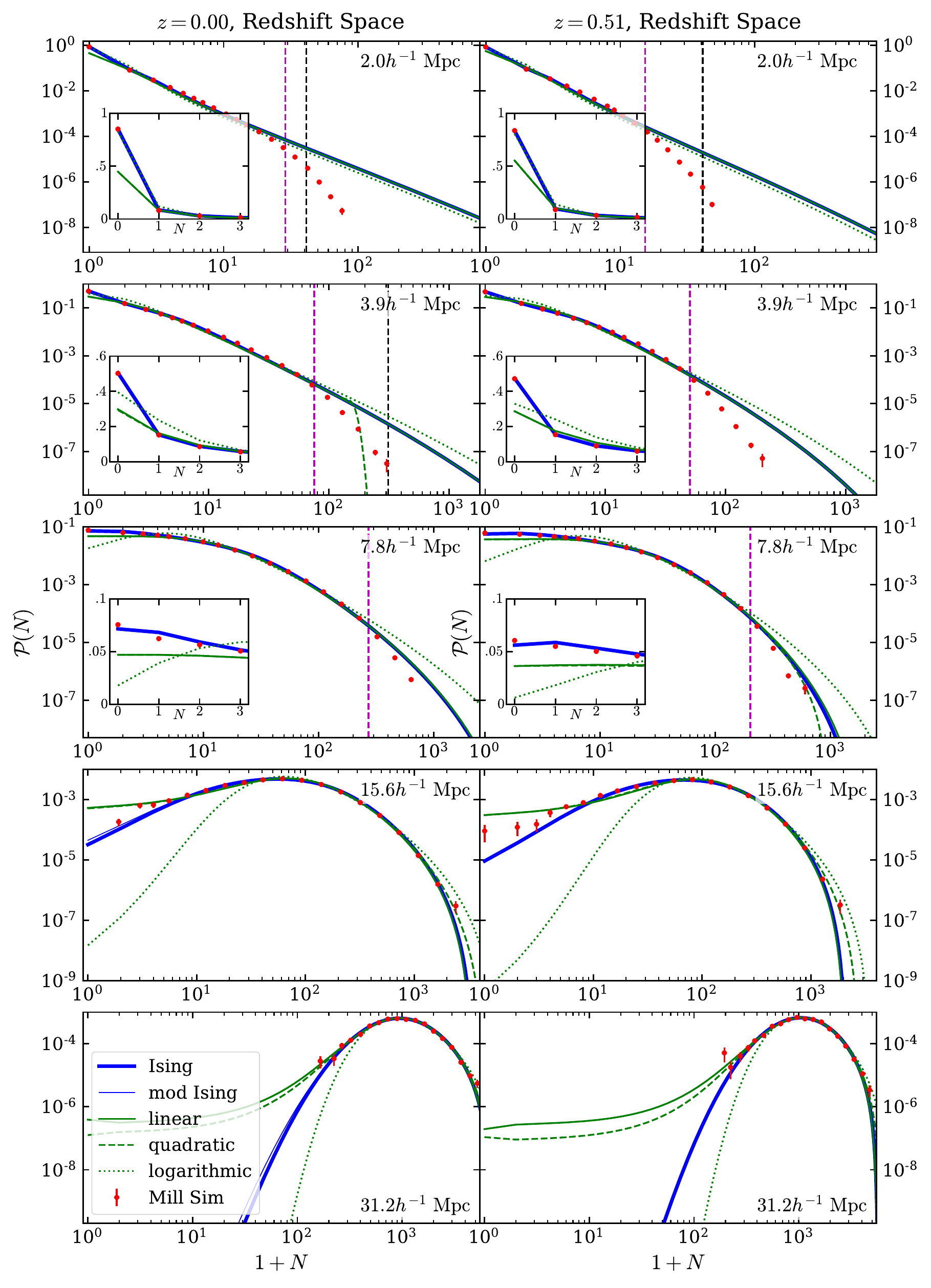}
\caption{Results of fitting bias models to Millennium Simulation galaxy distributions in redshift space at $z = 0$ (left panels) and $z = 0.51$ (right panels). Each row shows the probability distribution $\mathcal{P}(N)$ for a specified cubical cell size. Red points show the probabilities observed in the redshift-space simulation catalogs (see Section~\ref{sec:zsp});  thick blue and green curves show the predictions of five galaxy bias models. Insets show the low-$N$ distributions. Dashed black lines mark the values of $N_\mathrm{cut}$ adopted in Equations~\ref{eq:bigeq_unmod}--\ref{eq:defI2cut}; dashed magenta lines mark the values of $N_\mathrm{exp}$ estimated in Section~\ref{sec:disc} and displayed in Fig.~\ref{fig:Acutoff}.
}
\label{fig:zsp1}
\end{figure*}

\begin{figure*}
\leavevmode\epsfxsize=16cm\epsfbox{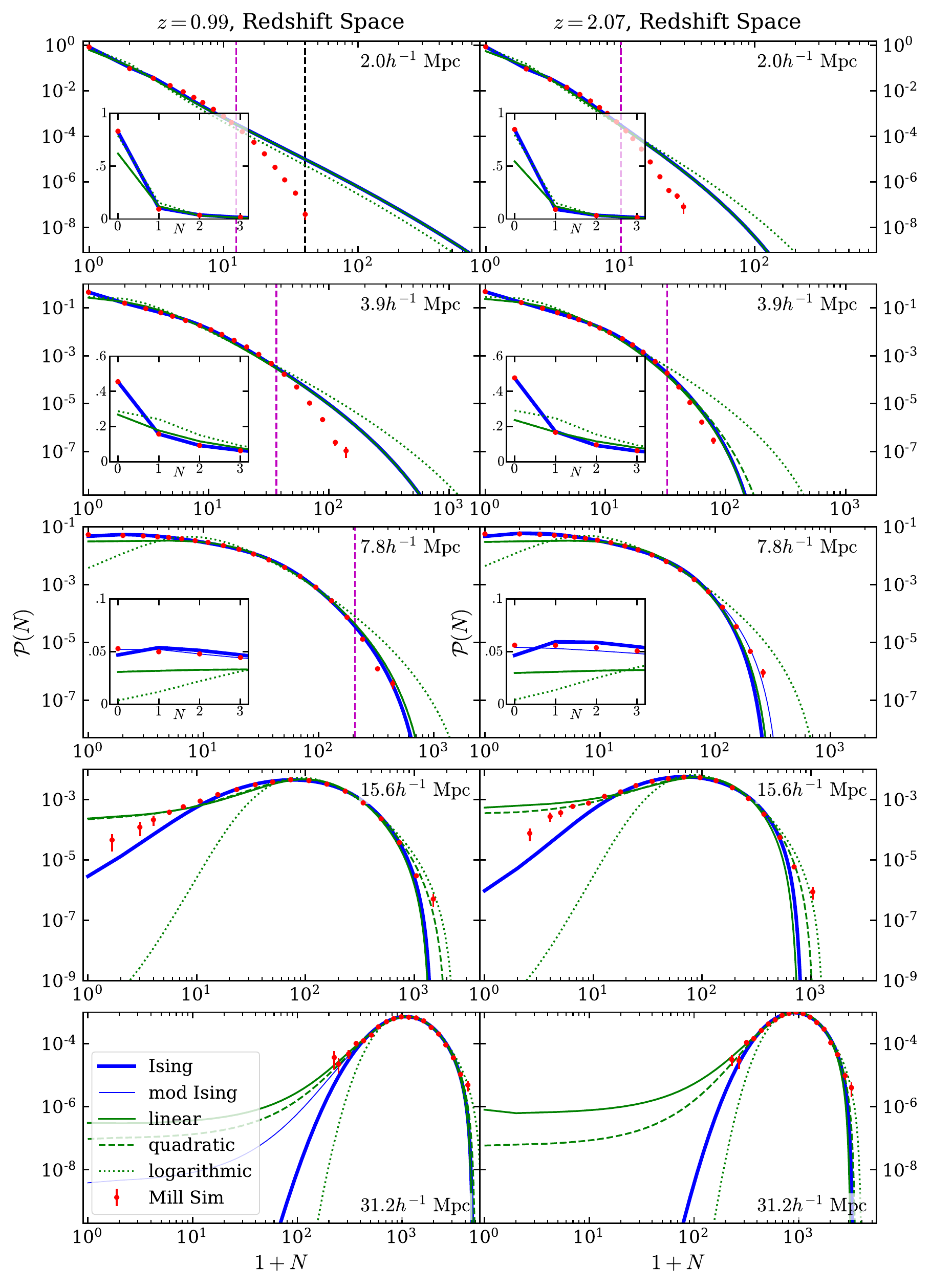}
\caption{Results of fitting bias models to Millennium Simulation galaxy distributions in redshift space at $z = 0.99$ (left panels) and $z = 2.07$ (right panels). See caption of Fig.~\ref{fig:zsp1} for further description.}
\label{fig:zsp2}
\end{figure*}

Nevertheless, before applying the $\mathcal{P}(N)$ model of Section~\ref{sec:PNmethod} to survey data, we must verify that our method can discriminate between bias models even in the presence of redshift-space distortion. To do so we consider the same Millennium Simulation catalogs as before (described at the start of Section~\ref{sec:PN} and employed in Section~\ref{sec:PNvalid}). Making use of the plane-parallel approximation, we perturb the $z$-coordinate of each galaxy by $v_z/100$, where $v_z$ is the galaxy's proper motion within the simulation (the factor of 100 is due to velocity and distance units being km/s and $h^{-1}$Mpc, respectively). We then repeat the fitting procedure of Section~\ref{sec:PNvalid} for these redshift-space-distorted galaxy catalogs; the resulting $\chi^2_\nu$-values appear in Fig.~\ref{fig:chi2dof} (as semi-transparent curves), and the fits themselves appear in Figs.~\ref{fig:zsp1} and \ref{fig:zsp2}.

The $\chi^2$ values plotted in Fig.~\ref{fig:chi2dof} show that, as expected, there is virtually no difference above $\sim 5h^{-1}$-Mpc scales in the quality of the Ising model fits in redshift and real space. (The modified Ising and linear/quadratic models seem to do worse in redshift space at these scales.) We expect to see the greatest distortions at smaller scales, where Finger-of-God effects become dominant. In agreement with this expectation, we note that the Ising-model fit at small scales in redshift space is significantly worse than in real space. Nevertheless, the Ising fits are still significantly better (by over an order of magnitude at the smallest scales) than those of the linear and quadratic models.

Figs.~\ref{fig:zsp1} and \ref{fig:zsp2} clarify the reason that none of the models yields a particularly good fit at small scales -- namely, there exists a steep dropoff in probability at large densities, which none of the models can reproduce. This dropoff is unsurprising given that Finger-of-God distortions ``smear'' dense clusters in the radial direction but do not affect low-density regions (except perhaps by smearing nearby clusters into them). Thus the incidence of already-rare dense clusters falls steeply upon passage to redshift space.

In Section~\ref{sec:disc} we further analyse the effects of redshift-space distortion and consider the possibility of modeling them. However, the goal of this paper is simply to demonstrate that the Ising bias model describes the empirical galaxy distribution better than the reference models. At this point we have outlined a method for doing so using the distribution of galaxy counts (Section~\ref{sec:PNmethod}); we have shown that the method effectively discriminates between bias models in real space (Section~\ref{sec:PNvalid}); and we have now shown that the method effectively discriminates between models even in redshift space (Section~\ref{sec:zsp}). In particular, we note that if a particular bias model does \emph{not} fit the true galaxy distribution, that mismatch is likely to be apparent in both redshift space and real space. Hence in the following section we proceed to apply this method to test the fit of bias models to empirical data.

\section{Comparison to Observational Data}
\label{sec:data}
Although cosmological simulations can yield significant insight, the true test of any theory or model must be observation of the actual Universe. Therefore we now present tests of the Ising bias model using three galaxy catalogs covering different redshift regimes: the 6dFGS galaxies at $z \sim 0.04$ (Section~\ref{sec:6dFGS}), the SDSS Main Galaxy Sample at $z \sim 0.14$ (Section~\ref{sec:SDSS}), and the COSMOS2015 galaxies at $z \sim 1.3$ (Section~\ref{sec:COSMOS}). We note that all three of these analyses necessarily occur in redshift space.

\subsection{6dFGS Galaxies}
\begin{figure*}
\leavevmode\epsfxsize=18cm\epsfbox{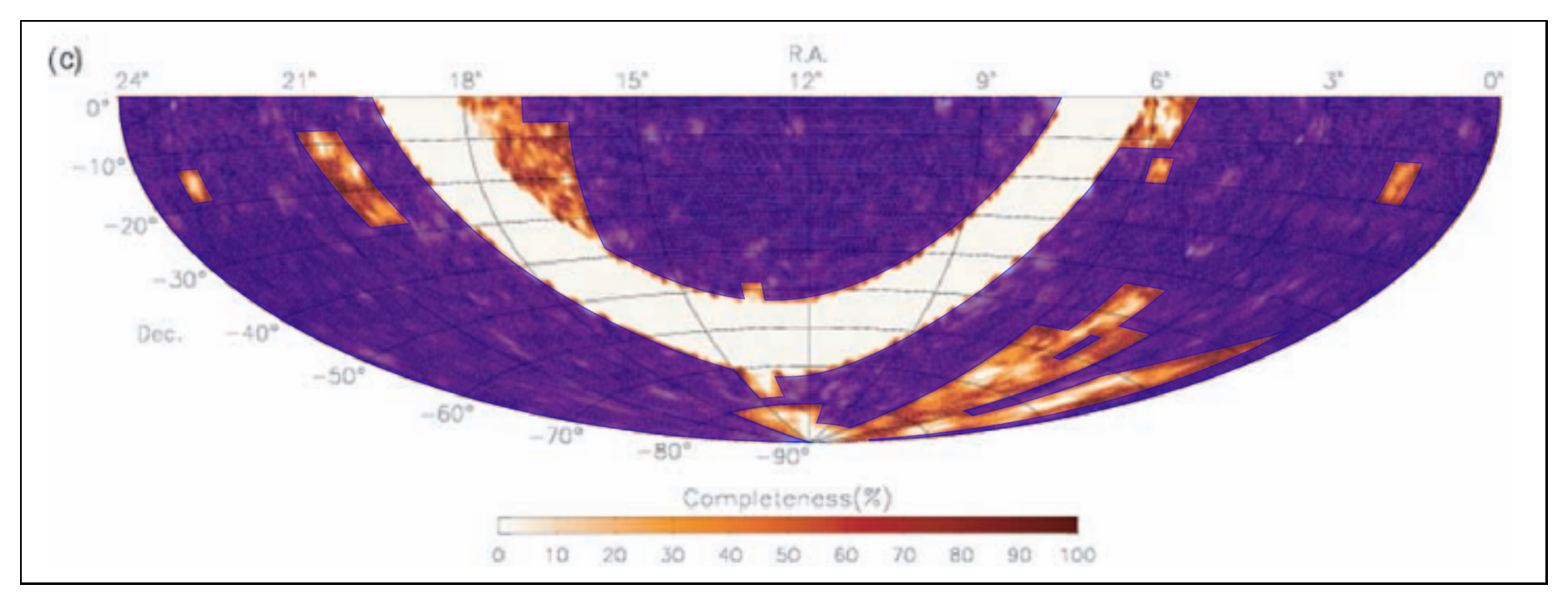}
\vspace{-0.7cm}
\caption{The area (in purple) from which we take our sample (Section~\ref{sec:6dFGS}) of 6dFGS galaxies, superimposed upon the 6dFGS completeness map from fig. 1(c) of \citet{Jones2009}.}
\label{fig:6dFGS_area}
\end{figure*}

The 6dF Galaxy Survey\footnote{http://www-wfau.roe.ac.uk/6dFGS} (6dFGS; \citealp{Jones2009}) utilized the 6dF multi-object spectrograph on the United Kingdom Schmidt Telescope (UKST) to obtain spectroscopic redshifts for over 110,000 galaxies; the survey covered $\sim 17,000$ square degrees on the sky, with a median galaxy redshift around $z=.05$. However, the survey completeness varies across the target region (see Fig.~\ref{fig:6dFGS_area}, the background of which is reproduced from \citealp{Jones2009}).

\begin{figure*}
\leavevmode\epsfxsize=16cm\epsfbox{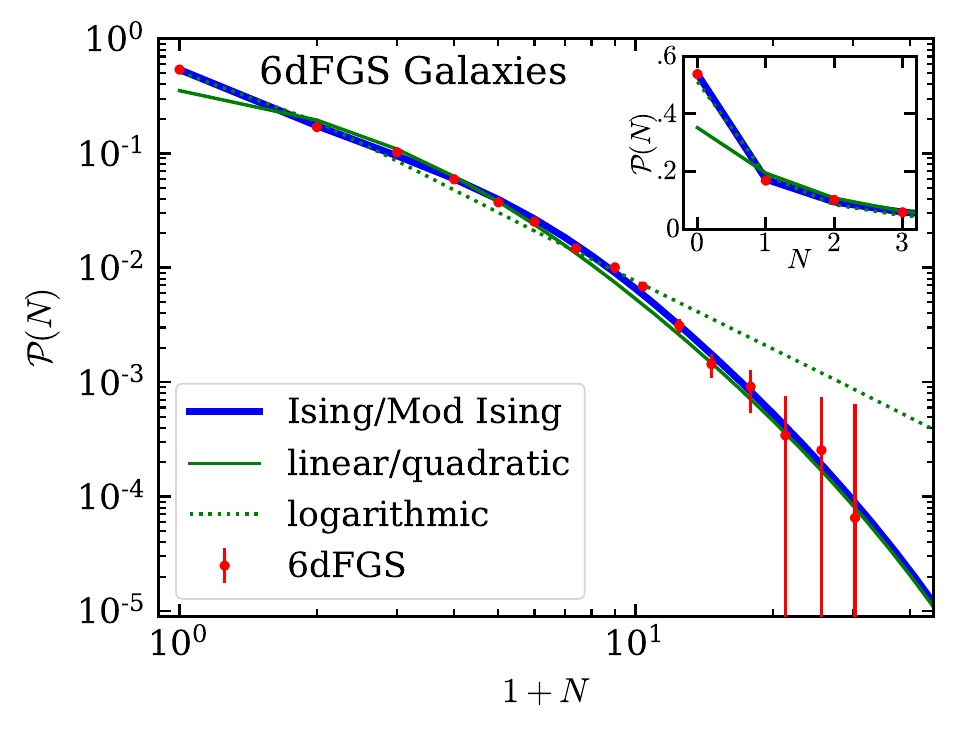}
\caption{Probability distribution of number of 6dFGS galaxies per survey cell, for $8h^{-1}$-Mpc cubical cells and the galaxy selection described in Section~\ref{sec:6dFGS}. Red points show the empirical results; blue and green curves show the predictions of various galaxy bias models. (The best-fitting modified Ising and quadratic models are in this case identical to the best-fitting Ising and linear models, respectively.)}
\label{fig:6dFGS_PN}
\end{figure*}

To build a uniform sample from this data set, we first exclude certain large regions containing areas of low completeness; after excluding these regions about 14,900 square degrees remain (shaded purple in Fig.~\ref{fig:6dFGS_area}). This remaining area, however, still suffers from significant completeness variations. By random sampling the electronic version of fig.~1(c) of \citet{Jones2009}, we estimate the standard deviation of completeness to be 0.18 of the mean completeness; we thus add (in quadrature) an uncertainty of 0.2 to the estimated counts variability from Appendix~\ref{app:PNerrs}. (It is for this reason that the error bars in Fig.~\ref{fig:6dFGS_PN} are noticeably larger than those in Figs.~\ref{fig:SDSS_PN} and \ref{fig:COSMOS_PN}.)
\label{sec:6dFGS}

We now select only galaxies in the redshift range $0.01 \le z \le 0.054$ (the full sample's median redshift being 0.053), requiring redshift quality flags $Q=3$ or 4. Since the survey's limiting magnitude is $K \le 12.65$ we conservatively require a $K$-band magnitude less than 11.90 for a galaxy at our upper redshift limit $z=0.054$ (scaled appropriately at lower redshifts), obtaining a volume-limited sample with $M_K \le -23.43 + 5 \log h$. Our final 6dFGS sample comprises about 19,000 galaxies with a median redshift of 0.0445.

We next count the galaxies in three-dimensional cubical cells with $8h^{-1}$Mpc-sides. Since some of these cells extend outside the volume of our sample, we discard any cells for which less than 90 percent is contained within the sample volume. We apply a simple fractional correction if the percentage of the cell extending outside the volume is between 0 and 10 percent. We can then determine the galaxy counts in each cell, obtaining 10,233 cells with an average of $\overline{N}=1.48$ galaxies per cell. Finally, we determine probabilities $\mathcal{P}(N)$ and fit the bias models as in Section~\ref{sec:PN}.

The results appear in Fig.~\ref{fig:6dFGS_PN}, with the best-fitting parameter values and reduced $\chi^2$-values in Table~\ref{tab:obs_PN}. In this case (and as previously noted for fits in redshift space) the quadratic model provides no improvement over the linear, nor the modified Ising over the Ising. The linear and quadratic bias models (due to their unphysical predictions) fail to fit the data at low densities, whereas the logarithmic model fits well at low densities but fails at moderate and (especially) high densities. The value of $\chi^2_\nu$ for the Ising model is almost two orders of magnitude lower than that for the linear and quadratic models. The disparity is not as great for the logarithmic model, but the logarithmic model obtains this result by sacrificing the greatest strength of the linear/quadratic models, namely, the fact that they fit the data well at the high densities which dominate the standard power-spectrum-based analysis.

We thus conclude that for this low-redshift sample, the Ising galaxy bias model provides a significantly superior description of the galaxy counts than do the reference models.

\subsection{SDSS Main Galaxy Sample}
\label{sec:SDSS}
\begin{figure*}
\leavevmode\epsfxsize=18cm\epsfbox{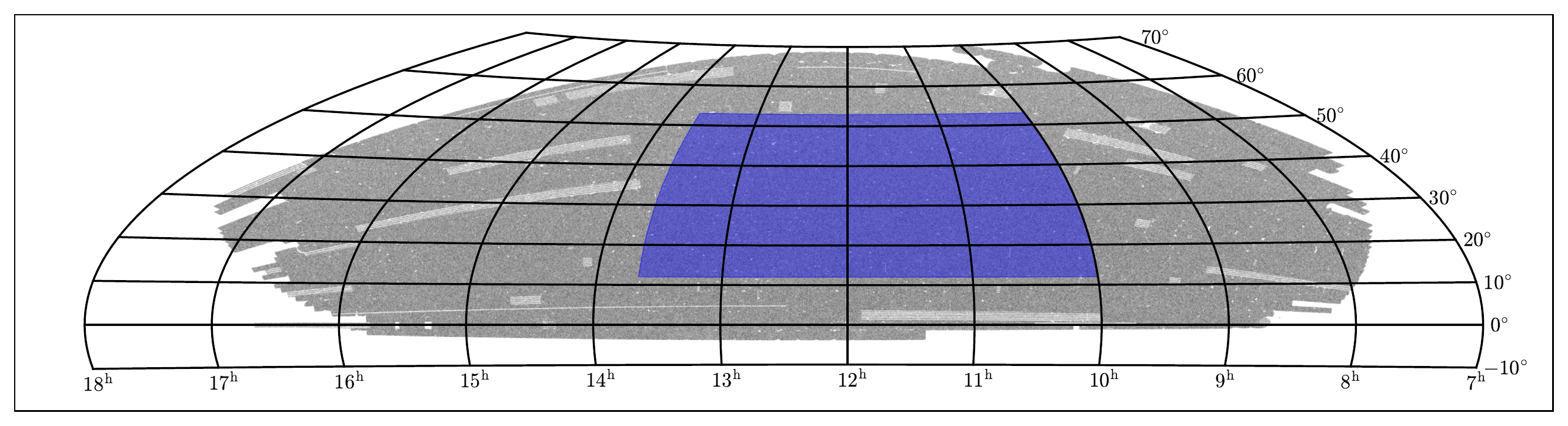}
\vspace{-0.7cm}
\caption{For the SDSS Main Galaxy Sample, we mark the area (in blue) from which we draw our test sample in Section~\ref{sec:SDSS}; we show the location of the SDSS MGS galaxies in grey.}
\label{fig:SDSS_area}
\end{figure*}

The Sloan Digital Sky Survey (SDSS; \citealp{York2000}) data release seven (DR7; \citealp{Abazajian2009}) includes the Main Galaxy Sample (MGS), a flux-limited redshift survey. For our analysis of the MGS, we obtain our data from the NYU-VAGC\footnote{http://sdss.physics.nyu.edu/vagc/lss.html} (New York University-Value Added Galaxy Catalog), compiled as in \citet{Blanton2005}. We restrict ourselves to the North Galactic Cap. Following \citet{Ross2015}, we employ the `safe0' catalog and apply the following cuts to obtain a roughly homogenous, volume-limited sample: $M_r < -21.2$, $g-r > 0.8$, $0.07 < z < 0.17$. Note that our upper redshift limit is lower than that of \citet{Ross2015} in order to avoid the turnover in number density noted in their fig. 2. The grey region of our Fig.~\ref{fig:SDSS_area} shows the location of these galaxies on the sky.

We bin these galaxies into three dimensional cubical cells of side length $16h^{-1}$Mpc in order to have a reasonable number of galaxies per cell. Since (small) areas of incomplete coverage are apparent in Fig.~\ref{fig:SDSS_area}, we use only cells with projected centres having right ascension $10^\mathrm{h}00^\mathrm{m} < \alpha < 13^\mathrm{h}40^\mathrm{m}$ and declination $12^\circ < \delta < 53^\circ$ (the region marked in blue on the figure), an area comprising about 1860 square degrees. We further (conservatively) retain only cells such that the redshifts of their centres satisfy $0.075 < z < 0.165$. These cuts yield 4450 cells with an average of $\overline{N}=2.13$ galaxies per cell and a median galaxy redshift of  $z=0.139$.Given these cells we can calculate probabilities $\mathcal{P}(N)$ and fit the bias models, following Section~\ref{sec:PN}.

\begin{figure*}
\leavevmode\epsfxsize=16cm\epsfbox{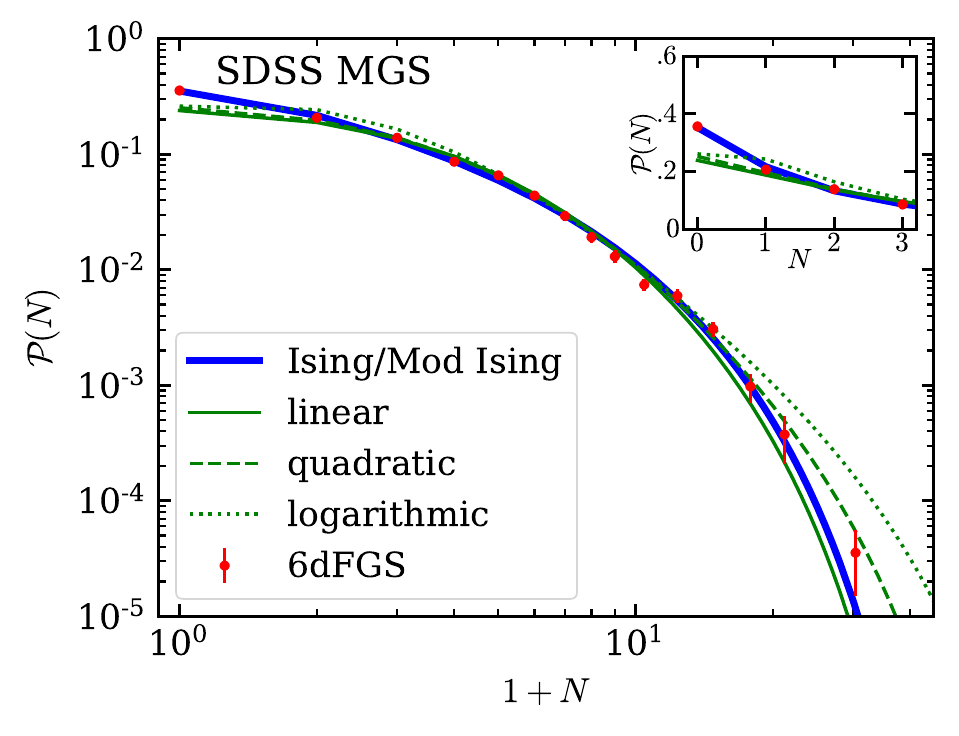}
\caption{Probability distribution of number of SDSS Main Galaxy Survey (MGS) galaxies per survey cell, for $16h^{-1}$-Mpc cubical cells and the galaxy selection described in Section~\ref{sec:SDSS}. Red points show the empirical results; blue and green curves show the predictions of various galaxy bias models. (The best-fitting modified Ising model is here indistinguishable from the best-fitting Ising model.)}
\label{fig:SDSS_PN}
\end{figure*}

Upon doing so, we obtain the results in Fig.~\ref{fig:SDSS_PN} (see Table~\ref{tab:obs_PN} for the best-fitting parameter- and reduced $\chi^2$-values). Again the (unmodified) Ising model provides a markedly superior fit, upon which the modified model fails to improve. The reference models (linear, quadratic, and now the logarithmic as well) fail to fit the data at low densities. The quadratic model provides the best alternative, but its value of $\chi^2_\nu$ is ten times higher than that of the Ising model. (Compare this result to the difference of two orders of magnitude for the 6dFGS sample, which probed length scales half as large as this one.)

We conclude that for this sample also, the Ising galaxy bias model provides a description of the galaxy counts which is superior to that yielded by the reference models.

\subsection{COSMOS Galaxies}
\label{sec:COSMOS}

\begin{figure}
\leavevmode\epsfxsize=8cm\epsfbox{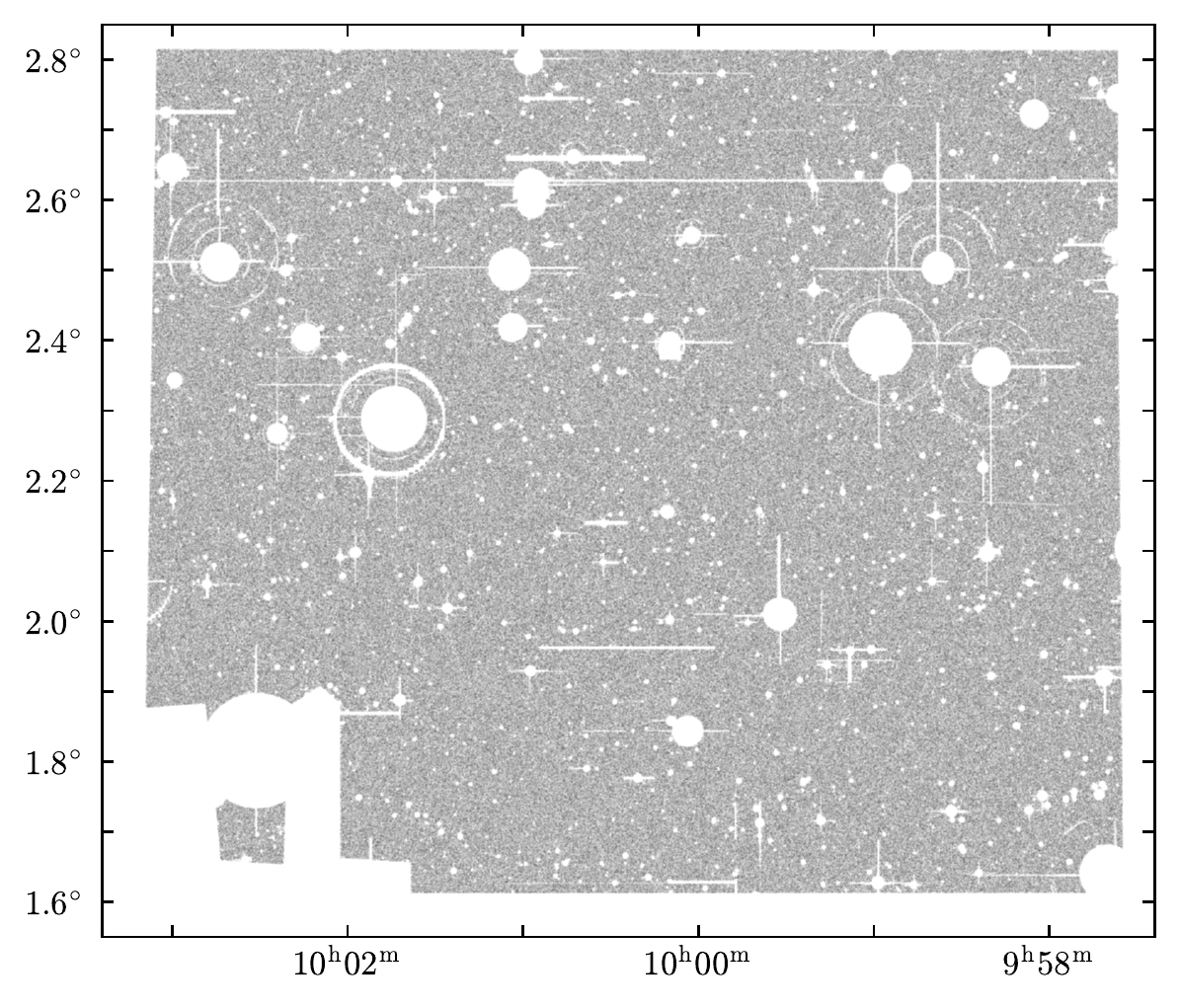}
\vspace{-0.2cm}
\caption{For analysis of COSMOS galaxies (Section~\ref{sec:COSMOS}), the area from which we take our sample.}
\label{fig:COSMOS_area}
\end{figure}

The COSMOS2015 galaxy catalog\footnote{ftp://ftp.iap.fr:/pub/from\_users/hjmcc/COSMOS2015} \citep{Laigle2016} contains over half a million galaxies with photometric redshifts up to $z \sim 6$. For our selection we use galaxies with UltraVISTA \citep{McCracken2012} data that fall within the COSMOS field: i.e., $\mathcal{A}^\mathrm{UVISTA} \cap \mathcal{A}^\mathrm{COSMOS}$ in the notation of \citet{Laigle2016}; we also exclude regions that are masked due to contamination by foreground stars, etc., for a resulting area of 1.38 square degrees, shown in Fig.~\ref{fig:COSMOS_area}.

This sample differs significantly from the 6dFGS and SDSS samples in that the COSMOS2015 redshifts are photometric rather than spectroscopic. As a result, the error inherent in the photometric redshift process smooths the galaxy distribution in the radial direction. The quoted accuracy of the COSMOS redshifts depends on the subsample under consideration, but we can obtain an  order-of-magnitude estimate of this smoothing effect by noting that the typical photometric redshift error $\sigma_{\Delta z/(1+z)}$ is on the order of 0.02 (see table 5 and fig. 14 of \citealp{Laigle2016}). At a median redshift of 1.263 this uncertainty yields a typical error $\sigma_{\Delta z} \sim 0.05$, which in the concordance cosmology is equivalent to a radial error around $60h^{-1}$Mpc. Hence, despite the fact that the COSMOS2015 photometric redshifts are in general exquisite, the error is nevertheless significantly larger than the cell sizes in this work.

\begin{figure}
\leavevmode\epsfxsize=9cm\epsfbox{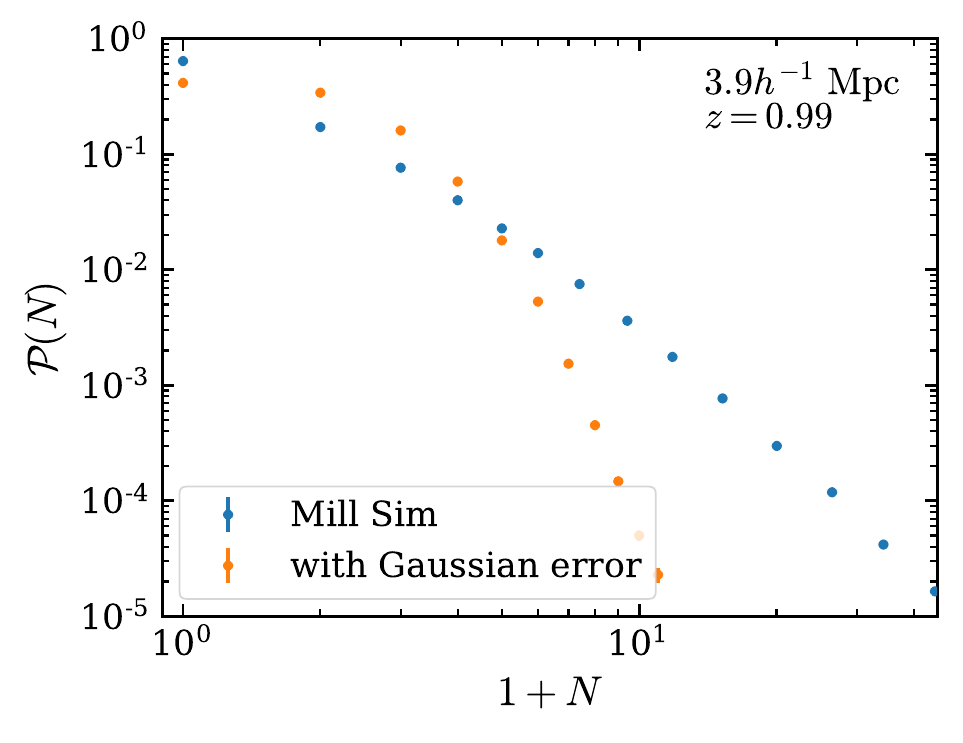}
\caption{The effect of adding a Gaussian perturbation (of size $60h^{-1}$Mpc) to one dimension of the Millennium Simulation galaxies (at $z = 0.99$) of Section~\ref{sec:PNvalid}. We note that the perturbed distribution (orange) contains more intermediate-density cells than the original distribution (blue -- compare to the left panel of the second row of Fig.~\ref{fig:MS_PN99}). The same effect is evident in the mismatch of the COSMOS2015 galaxies to realistic bias models in Fig.~\ref{fig:COSMOS_PN} as a result of the use of photometric redshifts in the COSMOS2015 catalog.}
\label{fig:MS_Gerr}
\end{figure}

We can estimate the effect of this photometric redshift error by returning to the Millennium Simulation galaxy catalogs and adding Gaussian errors (with standard deviation of $60h^{-1}$Mpc) to one dimension of the galaxy positions. To approximate our COSMOS analysis below, we chose the catalog with cells of side $3.9h^{-1}$Mpc at $z = 0.99$, and before adding the error we randomly select only enough galaxies to match the COSMOS number density. Fig.~\ref{fig:MS_Gerr} compares the resulting probability counts to those of the catalog without the added error, and it is clear that galaxies in the highest-density cells are more likely to scatter into their less-dense neighbors, thus reducing the number of high-density cells; whereas the lowest-density cells are more likely to gain galaxies than to lose them, and thus the number of low-density cells falls as well. The net effect is to reduce the incidence of both galaxy-rich and galaxy-poor cells relative to those of intermediate density.

Thus, the photometric redshift errors in the COSMOS catalog are significantly larger than the scale at which the Ising model is advantageous, and they signficantly change the shape of the probability distribution. Hence it would be surprising if the Ising model were to provide an accurate fit in this case. Indeed, a failure to fit the COSMOS2015 probabilities would confirm that the Ising model is falsifiable and thus meaningful, and for this reason we perform the analysis of the catalog as described below.

\begin{figure}
\leavevmode\epsfxsize=8cm\epsfbox{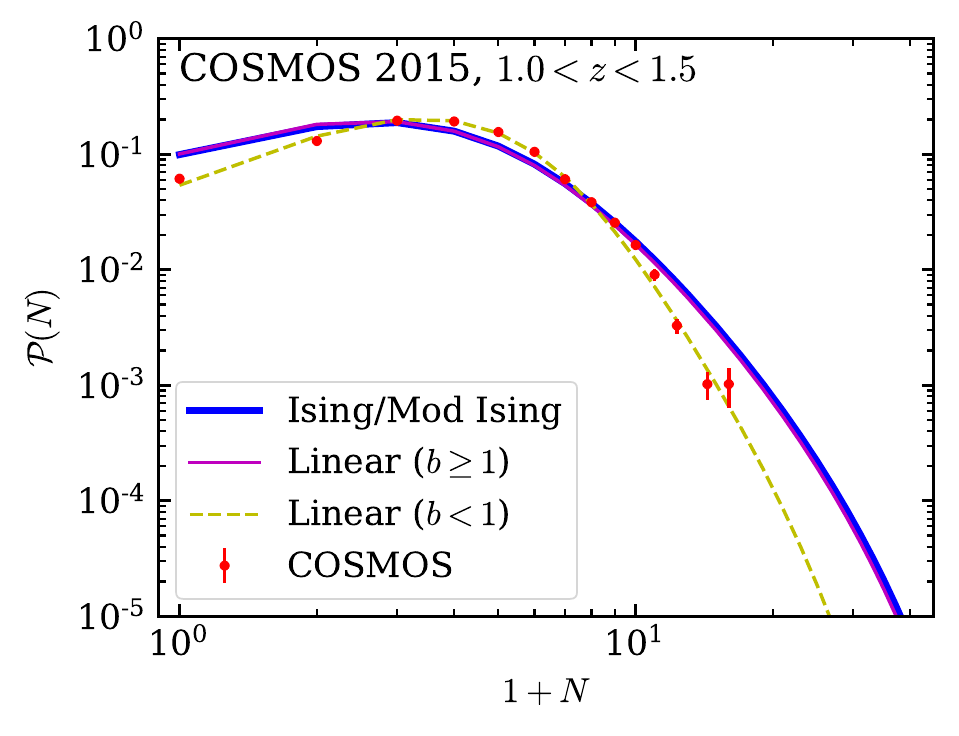}
\caption{Probability distribution of number of COSMOS2015 galaxies \citep{Laigle2016} per survey cell, for $6h^{-1}$-Mpc cubical cells and the galaxy selection described in Section~\ref{sec:COSMOS}. Red points show the empirical results in photometric-redshift space; the blue curve shows the best-fitting the Ising model. The magenta curve (barely distinguishable from the blue) shows the best-fitting linear bias model if we require a (realistic) bias not less than unity. The dashed yellow shows the best-fitting linear bias if we relax this restriction (in this case $b = 0.585$). The photometric-redshift uncertainties have distorted the curve to the point where none of the bias models (with realistic parameter values) provides a good fit.}
\label{fig:COSMOS_PN}
\end{figure}

From the COSMOS catalog we thus choose galaxies with (photometric) redshift $1.0 < z < 1.5$; since the $K$-band limiting magnitude is $24.0$, we impose this limit at $z=1.5$ and scale with redshift, thus requiring $M_K \le -21.27 + 5\log h$. Because of the small solid angle -- and because the number of galaxies in the catalog supports such a choice -- we employ smaller survey cells ($6h^{-1}$ Mpc per side) than for the previous data sets. As in Section~\ref{sec:6dFGS}, we reject any cell if more than 10 percent of its volume extends outside the sample volume, and we apply a completeness correction to each cell for which less than 10 percent of its volume extends outside the sample region. We thus obtain 6842 cells with $\overline{N}=3.44$ galaxies per cell. Given this selection, we then determine the probabilities and fit bias models as in Section~\ref{sec:PN}. The results appear in Fig.~\ref{fig:COSMOS_PN} and Table~\ref{tab:obs_PN}.

In this case, and as expected, the results diverge greatly from those obtained for the 6dFGS and SDSS samples. The Ising model cannot provide a good fit to the data. The other three models can do so only by adopting unphysically low values of the bias parameter ($b \approx 0.6$, implying that galaxies cluster \emph{less} strongly than dark matter). Note that imposition of the net-galaxy constraint, Equation~\ref{eq:constraint}, prevents the Ising model from following suit. If we, more realistically, require biases of at least unity, then none of the bias models can fit the observed probability distribution; indeed, all of the models are ``equally bad,'' with none of the $\chi^2_\nu$-values differing by more than 10 per cent from 30.

Upon closer consideration of this behavior, we see in Fig.~\ref{fig:COSMOS_PN} that the models overestimate the low- and high-$N$ probabilities and underestimate them near $\overline{N} = 3.44.$ Thus, the data show more intermediate-density cells -- and fewer high- and low-density cells -- than one would expect, precisely as suggested by the simulation results displayed in Fig.~\ref{fig:MS_Gerr}. (I.e., even with large photometric redshift errors, the data behave qualitatively as the simulations predict.) This result also explains the good fit of the dashed yellow fractional bias (``antibias'') curve in Fig.~\ref{fig:COSMOS_PN}: the highest-density cells appear to contain fewer galaxies than expected, and the lowest-density cells appear to contain more galaxies than expected. Or, in other words, the galaxies appear to cluster less strongly than dark matter because they have been smoothed out by the photometric redshift process. 

The smoothing effect of photometric redshift uncertainty is in some ways similar to that of redshift-space distortion (which we discuss further in Section~\ref{sec:disc}). However, photometric redshift error affects all galaxies (not only those in high-density regions), and it operates on much larger scales (in this case $\sim 60h^{-1}$Mpc rather than $\sim 5h^{-1}$Mpc); it thus prevents any of the bias models from reproducing the observed probability distribution in photometric-redshift space. We discuss prospects for dealing with these effects in the next section.

\begin{table*}
\begin{tabular}{ccccccccc}
\hline\\[-6mm]
Sample,                & \rule{-6mm}{0pt}Scale     & \rule{-2mm}{0pt}Bias     &                                                   &                       &Bias& Best-fitting             &\rule{0cm}{15pt} \\
Med $z$                & \rule{-6mm}{0pt}(Mpc$/h$) & \rule{-2mm}{0pt}Model    & Best-fitting Parameters                               & $\chi^2_\nu$          & Model & Parameters              & $\chi^2_\nu$\\ \hline\\[-4mm]
6dFGS                  & \rule{-6mm}{0pt}8.0       & \rule{-2mm}{0pt}Ising    &$b = 1.46$, $A_t = -.315$, $T = .321$              &$0.64$\rule{0cm}{12pt} & Linear& $b = 1.46$                & $53.5$\\
$z_\mathrm{md}=0.0445$ &                           & \rule{-2mm}{0pt}Mod Ising&$b = 1.46$, $A_t = -.315$, $T = .321$, $k = \infty$&$0.76$                 & Quad  & $b_1 = 1.46$, $b_2 = 0.00$& $57.6$\\
                       &                           &                          &                                                   &                       & Log   & $b = 2.17$                & $8.04$\\
SDSS MGS               & \rule{-6mm}{0pt}16        & \rule{-2mm}{0pt}Ising    &$b = 1.83$, $A_t = .002$, $T = .433$               &$0.82$ \rule{0cm}{15pt}& Linear& $b = 1.83$                & $10.8$\\
$z_\mathrm{md}=0.139$  &                           & \rule{-2mm}{0pt}Mod Ising&$b = 1.96$, $A_t = 0.00$, $T = .419$, $k = 2.66$   &$0.95$                 & Quad  & $b_1 = 1.83$, $b_2 = .362$& $8.82$\\
                       &                           &                          &                                                   &                       & Log   & $b = 1.68$                & $11.1$\\
COSMOS                 & \rule{-6mm}{0pt}6.0       & \rule{-2mm}{0pt}Ising    &$b = 1.03$, $A_t = -1.41$, $T = 0.00$              &$28.4$\rule{0cm}{15pt} & Linear& $b = .585$ (1.00)         & $1.49$ (27.2)\\
$z_\mathrm{md}=1.26$   &                           & \rule{-2mm}{0pt}Mod Ising&$b = 1.03$, $A_t = -1.41$, $T = 0.00$, $k = 1.84$  &$30.9$                 & Quad  & $b_1 = .584$, $b_2 = .009$& $1.62$\\
                       &                           &                          &                                                   &                       &       & (1.00, .026)              & $(29.2)$\\                    
                       &                           &                          &                                                   &                       & Log   & $b = .588$ (1.00)         & $3.07$ (27.2)\\
\hline
\end{tabular}
\caption{Results of fitting bias models to the probability distribution $\mathcal{P}(N)$ of the number of galaxies per survey cell, using the listed galaxy catalogs (described in Section~\ref{sec:6dFGS}--\ref{sec:COSMOS}). See graphical representation of survey areas in Figs.~\ref{fig:6dFGS_area}, \ref{fig:SDSS_area}, and \ref{fig:COSMOS_area}, and of fits in Figs.~\ref{fig:6dFGS_PN}, \ref{fig:SDSS_PN}, and \ref{fig:COSMOS_PN}. COSMOS entries in parentheses indicate the effect of requiring a realistic bias of at least unity.\label{tab:obs_PN}}
\end{table*}

\section{Discussion}
\label{sec:disc}

In this section, we begin by revisiting the impact of redshift-space distortion on the fits to $\mathcal{P}(N)$ (Figs.~\ref{fig:zsp1} and \ref{fig:zsp2}). The large-scale distortion is the result of  coherent infall into overdensities (and out of underdensities), producing an apparent flattening along the line of sight of large-scale overdensities and a corresponding elongation of large-scale underdensities. The effect is evident in our results only at the largest scales, where we see a slight increase in the probability of low-density cells, the result of elongation (and resulting increase in volume) of voids. One could model this effect as a predictable change of variables, beginning with the observed volume element $d^3x \rightarrow d^3x_z$ \citep{Kaiser1987}. The result would be an altered matter probability distribution $\mathcal{P}(A)$, with a corresponding impact on $\mathcal{P}(N)$ according to Equation~\ref{eq:probs}.

The Finger-of-God effect, by contrast, is apparent on intermediate and smaller scales and produces the sharp probability dropoff observed in the upper rows of Figs.~\ref{fig:zsp1} and \ref{fig:zsp2}. The thermal motion within virialized clusters causes this distortion, spreading out the most massive clusters along the line of sight in redshift space and thus reducing the incidence of extremely overdense cells. (The moderate-density cells accordingly become more probable, an increase barely noticeable on the logarithmic scale of Figs.~\ref{fig:zsp1} and \ref{fig:zsp2}.) This distortion represents the convolution, in one dimension, of the spatial density distribution with a density-dependent velocity distribution.

In order to properly account for this effect, one could, in the Press-Schechter formalism, integrate upward from the initial density corresponding to virialization at the redshift under consideration. We plan to undertake such work in the future; here, however, we can at least provide a qualitative and phenomenological description of the effect, as follows.

Considering the upper three rows of Figs.~\ref{fig:zsp1} and \ref{fig:zsp2}, we note that the drop in probability appears to be linear (in the log-log scale of the plots -- thus exponential in $\ln(1+N)$) after a certain point. By extrapolating this linear portion to the left, we can determine where it intersects the real-space probability and thus estimate the value $N_\mathrm{exp}$ at which this exponential cutoff becomes important. (Dashed magenta lines mark $N_\mathrm{exp}$ in Figs.~\ref{fig:zsp1} and \ref{fig:zsp2}.) Survey cells with $N \ll N_\mathrm{exp}$ galaxies do not contain a significant number of virialized clusters, whereas cells with $N \gg N_\mathrm{exp}$ contain enough such virialized clusters to affect the distribution in redshift space.

Hence at any given scale, we expect $N_\mathrm{exp}$ to represent a density (after smoothing) which corresponds to significant intra-cellular virialization. Furthermore, to quantify ``significance,'' one could expect $N_\mathrm{exp}$ to depend on the initial overdensity $\delta_\mathrm{init}$ normalized by the typical fluctuation depth $\sigma_\mathrm{init}$, or, as evolved linearly forward, on $\delta_\mathrm{lin}/\sigma_\mathrm{lin}$ (i.e., $\nu$ in the Press-Schechter formalism). Recalling that $A = \ln(1+\delta)$ is a good proxy for $\delta_\mathrm{lin}$ \citep{CarronSzapudi2013}, we write $A_\mathrm{exp} \equiv \ln(N_\mathrm{exp}/\overline{N})$ and consider the quantity $A_\mathrm{exp}/\sigma_\mathrm{lin}$.

\begin{figure}
\leavevmode\epsfxsize=9cm\epsfbox{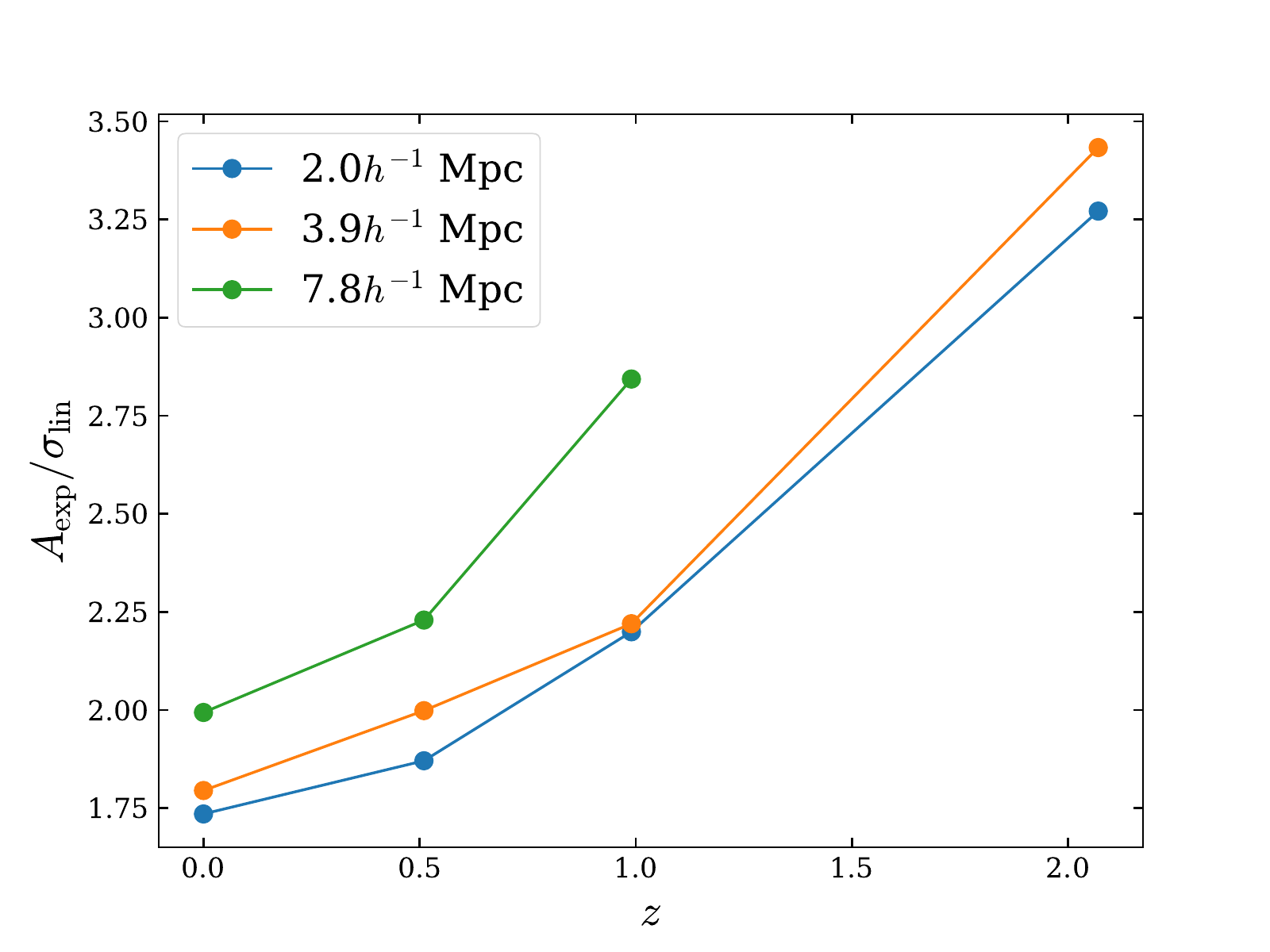}
\caption{Normalized log density $A_\mathrm{exp} = \ln(N_\mathrm{exp}/\overline{N})$ marking the location at which the exponential cutoff occurs in the redshift-space probability distribution $\mathcal{P}(N)$. Its increase with redshift signals the increasing rarity of virialized clusters; its increase with scale signals the increasing velocity dispersion necessary for a Finger-of-God to affect a survey cell.}
\label{fig:Acutoff}
\end{figure}

We can estimate $N_\mathrm{exp}$ at all four redshifts under consideration (see Figs.~\ref{fig:zsp1} and \ref{fig:zsp2}) for scales $2.0h^{-1}$ and $3.9h^{-1}$Mpc; at $7.8h^{-1}$Mpc, we can at least attempt to do so for $z = 0$, 0.51, and 0.99, although the estimates for this scale are likely to be somewhat inaccurate due to a paucity of available data points. Fig.~\ref{fig:Acutoff} displays the results as a function of redshift.

Inspecting Fig.~\ref{fig:Acutoff}, we note two trends. First, the quantity $A_\mathrm{exp}/\sigma_\mathrm{lin}$ increases with redshift, because at higher redshifts only the most extremely overdense regions have yet virialized, and thus only very overdense regions can contribute to the Finger-of-God effect. Second, $A_\mathrm{exp}/\sigma_\mathrm{lin}$ also increases with scale, both because larger scales ``dilute'' the effect (by smoothing over non-virialized neighboring regions) and also because the cells themselves are larger, requiring a more extreme velocity dispersion to ``smear'' clusters beyond the cell boundaries.

Finally, we note that the mechanism for modeling the effect of redshift-space distortion is conceptually similar to that  necessary for dealing with photometric-redshift errors, in that both require convolution of the density field (along one dimension) with a kernel that is at least roughly Gaussian. For redshift-space distortion, the kernel represents a velocity dispersion and is density-dependent, but for photometric-redshift errors, the kernel represents the redshift inaccuracy and is independent of density -- and thus perhaps easier to model.

This modeling is a priority for future work. Even without such modeling, one could apply the Ising bias model to two-dimensional survey results, or, alternatively, sort the galaxies into sufficiently wide redshift bins and interpret the results in each bin as a separate two-dimensional survey. In either case, the accuracy of the Ising model with respect to the reference models provides a better starting point for such procedures. In general, if the data support the introduction of additional parameters, one can augment the Ising model to increase its accuracy. The modified Ising model of Equations~\ref{eq:modIsing}--\ref{eq:defI2} provides an example of such potential modifications.
 
\section{Conclusions}
\label{sec:concl}
The use of galaxy surveys to constrain cosmology requires a model of galaxy bias; such models serve to map the underlying dark matter distribution to the observed number of galaxies. The Ising bias model is a physically motivated model that avoids the failures of the linear and quadratic models, namely, unphysical predictions in voids. This model also naturally incorporates stochasticity. \citet{Ising1} demonstrate (from Millennium Simulation galaxy counts) that the Ising model provides a significantly better description of the small-scale galaxy distribution than do the reference linear, quadratic, and logarithmic bias models.

In this work we apply the Ising model to empirical data. Doing so requires a method of testing bias models that does not require knowledge of the dark matter density in each cell. To formulate such a method, we  convolve the conditional probability distribution $\mathcal{P}(N|A)$ (derived from the bias models) with the underlying dark matter distribution $\mathcal{P}(A)$ obtaining for each model, a prediction of the distribution $\mathcal{P}(N)$ of galaxy counts. We then test this method using Millennium Simulation galaxy catalogs (in both real and redshift space); we find that the Ising model (vis-\`a-vis the reference models) consistently yields better fits to $\mathcal{P}(N)$ at small scales and comparable fits at large scales. Thus we have demonstrated the possibility of discriminating between various bias models using the observed probability distribution of galaxy counts.

We then evaluate the bias models by comparison to three galaxy data sets, namely, the 6dFGS galaxies, the SDSS Main Galaxy Sample, and the COSMOS2015 galaxies. The former two of these data sets use spectroscopic redshifts; the latter, photometric redshifts.

We find that for the spectroscopic-redshift data, the Ising model fits the observed probability distributions significantly better than the reference models (see Figs.~\ref{fig:SDSS_PN} and \ref{fig:6dFGS_PN}); in particular, the Ising model yields reduced $\chi^2$-values over ten times less than those of the reference models. We note that these results confirm the accuracy not only of the Ising bias model but also of the log-GEV dark matter distribution of \citet{ReppApdf}, which was assumed in the probability fits, with the qualifications outlined in Section~\ref{sec:PNmethod}.

For the COSMOS2015 catalog, only photometric redshifts are in general available. To investigate the effect of photometric redshift error, we apply a typical error to the Millennium Simulation catalog and obtain a significantly-altered probability distribution; we thus do not expect the bias models to yield a good fit. Congruent to these expectations, we find indeed that none of the models fits the data (if restricted to physically meaningful values) due to the extra smoothing induced by the photometric-redshift error. This negative result not only indicates the need for modeling the effect of this error but also demonstrates that our models (both the Ising bias model and the log-GEV dark matter distribution) are falsifiable given empirical probability counts. In particular, addition of a Gaussian perturbation of $60h^{-1}$Mpc to one dimension of the galaxies alters the distribution enough to make it incompatible with the Ising-bias/log-GEV matter model. Conversely, this fact emphasizes the significance of the fact that the Ising model provides a good fit to the spectroscopic redshift samples.

In conclusion, the Ising model of galaxy bias is physically motivated; it requires only two free parameters (being thus comparable to the quadratic model); it produces reasonable results in both low- and high-density regions; and it correctly models the probability distribution observed in spectroscopic galaxy surveys. It thus represents a simple solution to the problem of inferring dark matter statistics from galaxy counts. 

\section*{Acknowledgements}
The Millennium Simulation data bases used in this work and the web application providing online access to them were constructed as part of the activities of the German Astrophysical Virtual Observatory (GAVO). This work was supported by NASA Headquarters under the NASA Earth and Space Science Fellowship program -- ``Grant 80NSSC18K1081'' -- and AR gratefully acknowledges the support. IS acknowledges support from National Science Foundation (NSF) award 1616974. 

\begin{table*}
\begin{tabular}{cccccccc}
\hline\\[-6mm]
Scale & Bias  &                                          &                        &            & Bias     &  Best-fitting                  &\rule{0cm}{15pt} \\
(Mpc$/h$) & Model & Best-fitting Parameters                      & $\chi^2_\nu$           &\rule{1mm}{0cm}& Model &  Parameters                & $\chi^2_\nu$\\ \hline\\[-4mm]

$z = 0$:\hspace{3mm} & &                                      &                        &            &          &                            & \\
1.95  &Ising        & $b_I = 1.14$, $A_t =-.621$, $T = .214$           & $25.4$ &           &Linear      & $b = 1.14$                            & $4.40 \times 10^5$\\
         &Mod Ising& $b_I = 1.14$, $A_t=-.621$, $T=.215$, $k=82.7$ & $32.5$      &&Quadratic& $b_1 = 1.14$, $b_2 = 0.00$& $4.89 \times 10^5$\\
         &                &                                                                                   &                   &&Log          & $b = 1.15$                            & $2.10 \times 10^4$\\
3.91  &Ising        & $b_I = 1.22$, $A_t = -.895$, $T = .433$         & $34.0$ &         &Linear      & $b = 1.22$                            & $1.29 \times 10^4$\\
         &Mod Ising& $b_I = 1.22$, $A_t = -.895$, $T= .433$, $k=\infty$ & $40.1$&&Quadratic& $b_1 = 1.22$, $b_2 = 0.00$& $1.39 \times 10^4$\\
         &                &                                                                                   &              &    &Log          & $b = 1.32$                            & $4630$\\
7.81  &Ising        & $b_I = 1.19$, $A_t = -1.13$, $T = .471$           & $25.1$ &         &Linear      & $b = 1.21$                            & $70.7$\\
         &Mod Ising& $b_I = 1.19$, $A_t= -1.13$, $T=.471$, $k=\infty$ & $28.4$ & &Quadratic& $b_1 = 1.21$, $b_2 = 0.00$& $74.8$\\
         &                &                                                                                   &              &   &Log          & $b = 1.28$                            & $606$\\
15.6  &Ising        & $b_I = 1.15$, $A_t = -1.29$, $T = .568$           & $3.65$ &      &Linear      & $b = 1.15$                            & $2.79$\\
         &Mod Ising& $b_I = 1.23$, $A_t=-1.69$, $T=.442$, $k=.280$ & $2.20$ & &Quadratic& $b_1 = 1.17$, $b_2 = .042$& $2.38$\\
         &                &                                                                                   &              &   &Log          & $b = 1.22$                            & $30.5$\\
31.3  &Ising        & $b_I = 1.09$, $A_t = -1.42$, $T = .567$           & $0.64$ &     &Linear      & $b = 1.15$                            & $0.58$\\
         &Mod Ising& $b_I = 1.16$, $A_t=-1.30$, $T=.116$, $k=.024$ & $0.59$ &  &Quadratic& $b_1 = 1.14$, $b_2 = -.016$& $0.61$\\
         &                &                                                                                   &              &   &Log          & $b = 1.17$                            & $1.55$\\
$z = 0.51$:\hspace{3mm} & &                                      &                        &            &          &                            & \\
1.95  &Ising        & $b_I = 1.32$, $A_t =-.240$, $T = .342$           & $14.4$ &           &Linear      & $b = 1.32$                            & $8.71 \times 10^5$\\
         &Mod Ising& $b_I = 1.32$, $A_t=-.240$, $T=.342$, $k=75.5$ & $17.9$      &&Quadratic& $b_1 = 1.32$, $b_2 = 0.00$& $9.58 \times 10^5$\\
         &                &                                                                                   &                   &&Log          & $b = 1.22$                            & $2.91 \times 10^4$\\
3.91  &Ising        & $b_I = 1.32$, $A_t = -.565$, $T = .421$         & $39.5$ &         &Linear      & $b = 1.32$                            & $1.24 \times 10^4$\\
         &Mod Ising& $b_I = 1.40$, $A_t = -.954$, $T= .433$, $k=.466$ & $8.12$&&Quadratic& $b_1 = 1.32$, $b_2 = 0.00$& $1.33 \times 10^4$\\
         &                &                                                                                   &              &    &Log          & $b = 1.39$                            & $7010$\\
7.81  &Ising        & $b_I = 1.23$, $A_t = -.827$, $T = .440$           & $27.0$ &         &Linear      & $b = 1.32$                            & $77.9$\\
         &Mod Ising& $b_I = 1.36$, $A_t= -1.35$, $T=.468$, $k=.314$ & $5.00$ & &Quadratic& $b_1 = 1.32$, $b_2 = 0.00$& $82.4$\\
         &                &                                                                                   &              &   &Log          & $b = 1.35$                            & $715$\\
15.6  &Ising        & $b_I = 1.18$, $A_t = -.978$, $T = .499$           & $4.05$ &      &Linear      & $b = 1.24$                            & $3.70$\\
         &Mod Ising& $b_I = 1.27$, $A_t=-1.33$, $T=.447$, $k=.354$ & $2.69$ & &Quadratic& $b_1 = 1.25$, $b_2 = .038$& $3.60$\\
         &                &                                                                                   &              &   &Log          & $b = 1.28$                            & $43.6$\\
31.3  &Ising        & $b_I = 1.12$, $A_t = -1.07$, $T = .486$           & $0.88$ &     &Linear      & $b = 1.23$                            & $0.84$\\
         &Mod Ising& $b_I = 1.21$, $A_t=-1.41$, $T=.258$, $k=.044$ & $0.88$ &  &Quadratic& $b_1 = 1.22$, $b_2 = -.041$& $0.88$\\
         &                &                                                                                   &              &   &Log          & $b = 1.24$                            & $2.19$\\         
$z = 0.99$:\hspace{3mm} & &                                      &                        &            &          &                            & \\
1.95  &Ising        & $b_I = 1.52$, $A_t =.048$, $T = .381$           & $9.40$ &           &Linear      & $b = 1.52$                            & $9.77 \times 10^5$\\
         &Mod Ising& $b_I = 1.52$, $A_t=.0490$, $T=.382$, $k=73.0$ & $11.4$      &&Quadratic& $b_1 = 1.52$, $b_2 = 0.00$& $1.07 \times 10^6$\\
         &                &                                                                                   &                   &&Log          & $b = 1.28$                            & $3.63 \times 10^4$\\
3.91  &Ising        & $b_I = 1.44$, $A_t = -.309$, $T = .406$         & $63.0$ &         &Linear      & $b = 1.44$                            & $1.17 \times 10^4$\\
         &Mod Ising& $b_I = 1.44$, $A_t = -.309$, $T= .406$, $k=\infty$ & $72.0$&&Quadratic& $b_1 = 1.44$, $b_2 = 0.00$& $1.24 \times 10^4$\\
         &                &                                                                                   &              &    &Log          & $b = 1.44$                            & $9430$\\
7.81  &Ising        & $b_I = 1.30$, $A_t = -.572$, $T = .415$           & $31.5$ &         &Linear      & $b = 1.46$                            & $100.$\\
         &Mod Ising& $b_I = 1.48$, $A_t= -1.05$, $T=.489$, $k=.409$ & $4.19$ & &Quadratic& $b_1 = 1.46$, $b_2 = 0.00$& $106$\\
         &                &                                                                                   &              &   &Log          & $b = 1.41$                            & $845$\\
15.6  &Ising        & $b_I = 1.24$, $A_t = -.714$, $T = .465$           & $4.77$ &      &Linear      & $b = 1.35$                            & $4.43$\\
         &Mod Ising& $b_I = 1.38$, $A_t=-1.13$, $T=.459$, $k=.378$ & $3.16$ & &Quadratic& $b_1 = 1.36$, $b_2 = .055$& $4.26$\\
         &                &                                                                                   &              &   &Log          & $b = 1.37$                            & $50.0$\\
31.3  &Ising        & $b_I = 1.22$, $A_t = -.806$, $T = .525$           & $0.86$ &     &Linear      & $b = 1.34$                            & $0.71$\\
         &Mod Ising& $b_I = 1.34$, $A_t=-.885$, $T=.069$, $k=.003$ & $0.80$ &  &Quadratic& $b_1 = 1.34$, $b_2 = -.015$& $0.75$\\
         &                &                                                                                   &              &   &Log          & $b = 1.35$                            & $2.31$\\            
              
\hline
\end{tabular}
\caption{Results of fitting bias models to the probability $\mathcal{P}(N)$ of the number of galaxies per survey cell, using Millennium Simulation galaxy catalogs. Shown are the best-fitting values (and corresponding reduced $\chi^2$ values, for $N \le N_\mathrm{cut}$) for five bias models at various smoothing scales and redshifts; see graphical representations in Figs.~\ref{fig:chi2dof} and \ref{fig:MS_PN0}--\ref{fig:MS_PN2}. Values of $k=\infty$ indicate that the fitting algorithm was unable to find a modified Ising model with a $\chi^2$ less than that of the best-fitting Ising model. (Table continues on the following page.) \label{tab:MS_PN}}
\end{table*}
\begin{table*}
\begin{tabular}{cccccccc}
\hline\\[-6mm]
Scale & Bias  &                                          &                        &            & Bias     &  Best-fitting                  &\rule{0cm}{15pt} \\
(Mpc$/h$) & Model & Best-fitting Parameters                      & $\chi^2_\nu$           &\rule{1mm}{0cm}& Model &  Parameters                & $\chi^2_\nu$\\ \hline\\[-4mm]         
$z = 2.07$:\hspace{3mm} & &                                      &                        &            &          &                            & \\
1.95  &Ising        & $b_I = 2.22$, $A_t =.512$, $T = .393$           & $77.0$ &           &Linear      & $b = 2.15$                            & $1.13 \times 10^6$\\
         &Mod Ising& $b_I = 2.40$, $A_t=.162$, $T=.499$, $k=1.06$ & $23.4$      &&Quadratic& $b_1 = 2.15$, $b_2 = 0.00$& $1.22 \times 10^6$\\
         &                &                                                                                   &                   &&Log          & $b = 1.46$                            & $4.58 \times 10^4$\\
3.91  &Ising        & $b_I = 1.89$, $A_t = .113$, $T = .365$         & $127$ &         &Linear      & $b = 1.90$                            & $1.64 \times 10^4$\\
         &Mod Ising& $b_I = 1.89$, $A_t = .113$, $T= .365$, $k=\infty$ & $145$&&Quadratic& $b_1 = 1.90$, $b_2 = 0.00$& $1.74 \times 10^4$\\
         &                &                                                                                   &              &    &Log          & $b = 1.71$                            & $1.26e4$\\
7.81  &Ising        & $b_I = 1.61$, $A_t = -.146$, $T = .363$           & $51.3$ &         &Linear      & $b = 1.85$                            & $219$\\
         &Mod Ising& $b_I = 2.10$, $A_t= -.671$, $T=.595$, $k=.537$ & $8.06$ & &Quadratic& $b_1 = 1.93$, $b_2 = .193$& $214$\\
         &                &                                                                                   &              &   &Log          & $b = 1.73$                            & $921$\\
15.6  &Ising        & $b_I = 1.54$, $A_t = -.246$, $T = .413$           & $6.76$ &      &Linear      & $b = 1.69$                            & $12.0$\\
         &Mod Ising& $b_I = 2.20$, $A_t=-1.95$, $T=1.00$, $k=.261$ & $3.56$ & &Quadratic& $b_1 = 1.75$, $b_2 = .377$& $7.16$\\
         &                &                                                                                   &              &   &Log          & $b = 1.71$                            & $58.7$\\
31.3  &Ising        & $b_I = 1.52$, $A_t = -.284$, $T = .450$           & $0.59$ &     &Linear      & $b = 1.72$                            & $0.53$\\
         &Mod Ising& $b_I = 1.52$, $A_t=-.284$, $T=.450$, $k=\infty$ & $0.68$ &  &Quadratic& $b_1 = 1.74$, $b_2 = .255$& $0.41$\\
         &                &                                                                                   &              &   &Log          & $b = 1.70$                            & $3.19$\\            
              
\hline
\end{tabular}
\contcaption{(see caption on previous page.)\label{tab:MS_PN}}
\end{table*}

\bibliographystyle{astron}
\bibliography{Thesis_Proposal}

\appendix
\section{Derivation of Equations~15 and 18}
\label{app:cut}

We here demonstrate how to modify Equation~\ref{eq:constraint} to restrict its consideration to $A < A_\mathrm{cut}$. We assume that $A_\mathrm{cut} \gg A_t$, so that we are on the upper asymptote of the Ising function, where $\langle N \rangle_A = b_I \overline{N} e^A$; we also assume that $\langle N \rangle_A$ is high enough that the Poisson probability distribution $\mathrm{Pois}\left(N;\langle N \rangle_A\right)$ is approximately a Dirac delta function. Pursuant to these assumptions, we define
\begin{equation}
N_\mathrm{cut} = b_I \overline{N} \exp A_\mathrm{cut}.
\end{equation}
We also let $N_\mathrm{cells}$ be the number of cells in the galaxy survey.

We now let $N_\mathrm{cum}$ be the total number of galaxies in cells containing no more than $N_\mathrm{cut}$ galaxies, so that
\begin{align}
N_\mathrm{cum} & = N_\mathrm{cells} \sum_{N \le N_\mathrm{cut}} N\mathcal{P}(N)\label{eq:defNcum}\\
  & = N_\mathrm{cells} \sum_{N \le N_\mathrm{cut}} N \int dA\,\mathcal{P}(A) \mathcal{P}(N|A)\\
  & = N_\mathrm{cells} \sum_0^\infty N \int_{-\infty}^{A_\mathrm{cut}} dA\,\mathcal{P}(A) \mathcal{P}(N|A),
\end{align}
where the final step follows from our assumption that the distribution $\mathcal{P}(N|A)$ is approximately $\delta_D\left(N - \langle N \rangle_A \right)$.

Employing Equation~\ref{eq:final_FD}, we now obtain
\begin{align}
N_\mathrm{cum} & = N_\mathrm{cells} \int_{-\infty}^{A_\mathrm{cut}} dA\,\mathcal{P}(A) \langle N \rangle_A\label{eq:A5}\\
   & = N_\mathrm{cells} \int_{-\infty}^{A_\mathrm{cut}} dA\,\mathcal{P}(A) \frac{b_I \overline{N} e^A}{1+\exp\left(\frac{A_t-A}{T}\right)},
\end{align}
and thus by reference to Equation~\ref{eq:defNcum} we have
\begin{equation}
\sum_{N \le N_\mathrm{cut}} N\,n(N) = N_\mathrm{cells} \int_{-\infty}^{A_\mathrm{cut}} dA\,\mathcal{P}(A) \frac{b_I \overline{N} e^A}{1+\exp\left(\frac{A_t-A}{T}\right)},
\end{equation}
where $n(N) = N_\mathrm{cells}\mathcal{P}(N)$ is the number of survey cells containing $N$ galaxies. It follows that
\begin{equation}
b_I = \frac{\displaystyle\sum_{N=0}^{N_\mathrm{cut}} N \cdot n(N)/(N_\mathrm{cells}\overline{N})}{\displaystyle\int_{-\infty}^{A_\mathrm{cut}} dA\,\mathcal{P}(A) \frac{e^A}{1+\exp\left(\frac{A_t-A}{T}\right)}}.
\end{equation}
If we define, for later use,
\begin{align}
f & = \frac{1}{N_\mathrm{cells}\overline{N}}\sum_{N=0}^{N_\mathrm{cut}} N\cdot n(N)\label{eq:deffap}\\
I_1^\mathrm{cut} & = \int_{-\infty}^{A_\mathrm{cut}} dA\,\mathcal{P}(A) \frac{e^A}{1+\exp\left(\frac{A_t-A}{T}\right)},
\end{align}
we obtain
\begin{equation}
b_I = f/I_1^\mathrm{cut},
\end{equation}
which is Equation~\ref{eq:bigeq_unmod}.

The analysis for the modified Ising model (Equation~\ref{eq:modIsing}) is similar. In this case, from Equations~\ref{eq:defNcum}, \ref{eq:A5}, and \ref{eq:deffap}, we can write
\begin{align}
fN_\mathrm{tot} = {}& N_\mathrm{cells} \int_{-\infty}^{A_\mathrm{cut}} dA\,\mathcal{P}(A) \langle N \rangle_A\\
\begin{split}
    ={}& N_\mathrm{cells} b_I \overline{N} \int_{-\infty}^{A_\mathrm{cut}} dA\,\mathcal{P}(A) \frac{e^A}{1+\exp\left(\frac{A_t-A}{T}\right)}\\
         & + N_\mathrm{cells} \left(1-b_I\right) \overline{N} \int_{-\infty}^{A_\mathrm{cut}} dA\,\mathcal{P}(A) \exp \left( -k e^{-A} \right)
\end{split}\\
 f = {}& b_I I_1^\mathrm{cut} + \left(1 - b_I\right) I_2^\mathrm{cut},
\end{align}
where
\begin{equation}
I_2^\mathrm{cut} = \int_{-\infty}^{A_\mathrm{cut}} dA\,\mathcal{P}(A) \exp \left( -k e^{-A} \right).
\end{equation}
It follows that
\begin{equation}
b_I = \frac{f - I_2^\mathrm{cut}}{I_1^\mathrm{cut} - I_2^\mathrm{cut}},
\end{equation}
which is Equation~\ref{eq:bigeqmod}.

\section{Details on Fitting $\mathcal{P}(N)$}
\label{app:PNerrs}

In this appendix we provide additional details regarding our procedure for fitting bias models to the distribution of galaxy counts, whether obtained from simulations or observations.

\subsection{Errors on $\mathcal{P}(N)$}

A meaningful $\chi^2$ fit requires quantification of the error in the measured probabilities $\mathcal{P}(N)$, where $\mathcal{P}(N)$ is the probability of measuring $N$ galaxies in a survey cell.

The following provides a simple estimate, which however assumes vanishing covariance. As a result, some of the $\chi^2_\nu$-values in Tables~\ref{tab:obs_PN} and \ref{tab:MS_PN} are significantly lower than unity. Nevertheless, their relative values (and thus their ratios) provide a meaningful comparison between bias models.

To start, for any given whole number $N$, we imagine randomly selecting a survey cell; we then assign that cell a value $\mathcal{M} = 1$ if it contains $N$ galaxies and $\mathcal{M} = 0$ if it does not. It follows that the first two moments of $\mathcal{M}$ are
\begin{equation}
\langle \mathcal{M}^2 \rangle = \langle \mathcal{M} \rangle = \mathcal{P}(N),
\end{equation}
and the variance
\begin{equation}
\sigma_\mathcal{M}^2 = \mathcal{P}(N) - \mathcal{P}(N)^2.
\end{equation}
But if $N_\mathrm{tot}$ is the total number of cells in the survey, then we have $N_\mathrm{tot}$ measurements of $\mathcal{M}$. It follows that 
\begin{align}
\sigma_{\mathcal{P}(N)}^2 & = \sigma_{\langle M \rangle}^2 = \frac{\sigma_\mathcal{M}^2}{N_\mathrm{tot}} = \frac{\mathcal{P}(N) - \mathcal{P}(N)^2}{N_\mathrm{tot}}\\
   & = \frac{\mathcal{P}(N) \left( 1 - \mathcal{P}(N) \right)}{N_\mathrm{tot}},\label{eq:sigPN}
\end{align}
which is the prescription we use for the error on $\mathcal{P}(N)$.

Equation~\ref{eq:sigPN} requires modification when there is significant variation in survey completeness from cell to cell, as with the 6dFGS galaxies (Section~\ref{sec:6dFGS}). In general, given a whole number $N$, we will find $n_c$ cells (out of a total $N_\mathrm{tot}$ survey cells) which contain $N$ galaxies, so that $\mathcal{P}(N) = n_c/N_\mathrm{tot}$. With variable completeness, there are now two sources of error in $n_c$, namely, the stochastic variability modeled in Equation~\ref{eq:sigPN} and the completeness variability, which we parametrize as $\sigma_C = f_C \cdot n_c$. Then we can write
\begin{align}
\tilde{\sigma}_{\mathcal{P}(N)}^2 & = \sigma^2_{\mathcal{P}(N)} + \left(\frac{\partial \mathcal{P}(N)}{\partial n_c}\right)^2 \sigma_C^2 \\
   & = \sigma^2_{\mathcal{P}(N)} + \left(\frac{n_c f_C}{N_\mathrm{tot}} \right)^2.\label{eq:sigPNtild}
\end{align}
Based on sampling the shaded areas in Fig.~\ref{fig:6dFGS_area}, we estimate $f_C \sim 0.2$ and use this value in Equation~\ref{eq:sigPNtild} to obtain $\tilde{\sigma}_{\mathcal{P}(N)}$, which we use as the uncertainty in our probability values for Section~\ref{sec:6dFGS}.

\subsection{Binning Strategy}

It remains to detail our procedure for binning values of $N$ (i.e., galaxies per cell) to estimate probabilities. Since the (reduced) $\chi^2_\nu$ values are somewhat sensitive to the binning strategy, we use the same strategy for each experiment.

Thus, we begin with 20 bins in $N$-space. If the survey includes cells with no galaxies and cells with only one galaxy, we assign $N=0$ and $N=1$ to separate bins; if not, we ensure that the lowest bin contains only the lowest $N$-value. We then space the remainder of the bins logarithmically up to the maximum value of $N$ from the survey. Finally, we combine bins (starting with the lowest) as necessary to insure that no bin contains fewer than three cells.

For each bin we must also determine values for $\hat{N}$, $\hat{\mathcal{P}}(N)$, and $\hat{\sigma}^2_{\mathcal{P}(N)}$, where the circumflexes (hats) denote values estimated from the bins. We take $\hat{N} = \langle N \rangle_\mathrm{bin}$, the mean of the counts-in-cells values within the bin. Next, if $\Delta N$ is the (integer) width of the bin, then the value
\begin{equation}
\sum_{N \in \textrm{bin}} n_c(N)/N_\mathrm{tot} = \mathcal{P}\left(\lbrace N | N \in \textrm{bin} \rbrace\right),
\end{equation}
so that for any single $N$ in the bin we have
\begin{equation}
\hat{\mathcal{P}}(N) = \frac{\sum_{N \in \mathrm{bin}} n_c(N)}{N_\mathrm{tot} \cdot \Delta N}.
\end{equation}
It is this probability which we assign to $\hat{N}$. 

Likewise, for the measurement of the variance in the bin, we consider that we have $\Delta N$ independent measurements of $\hat{\mathcal{P}}(N)$ in that bin; thus we divide the values from Equation~\ref{eq:sigPN} (or \ref{eq:sigPNtild}) by $\Delta N$ to obtain $\hat{\sigma}^2_{\mathcal{P}(N)}$.

\label{lastpage}

\end{document}